\documentclass[12pt, draftclsnofoot, onecolumn]{IEEEtran}

\ifCLASSINFOpdf
  \usepackage[pdftex]{graphicx}
   \graphicspath{{..1/pdf/}{../jpeg/}}
   \DeclareGraphicsExtensions{.pdf,.jpeg,.png}
\else
   \usepackage[dvips]{graphicx}
   \graphicspath{{../eps/}}
   \DeclareGraphicsExtensions{.eps}
\fi

\usepackage{tikz}
\usetikzlibrary{calc,3d}
\usepackage[noadjust]{cite}
\usepackage{comment} 
\usepackage[cmex10]{amsmath}
\usepackage{amsfonts, amsmath, amsthm, amstext, amssymb, mathrsfs, graphicx, color, url}
\usepackage{subcaption}
\usepackage{array}
\usepackage{mdwmath} 
\usepackage{mdwtab}
\usepackage{algpseudocode}
\usepackage{algorithm}
\usepackage{url}
\usepackage{setspace}
\usepackage{footnote}
\usepackage[skip=0pt]{caption}
\usepackage{eqparbox}
\usepackage{fixltx2e}
\usepackage[utf8]{inputenc}
\usepackage{multicol}
\usepackage{mdframed}

\def\IB{{\mathbb B}}
\def\IC{{\mathbb C}}

\def\IE{{\mathbb E}}

\def\IR{{\mathbb R}}

\def\IZ{{\mathbb Z}}

\def\calA{{\mathcal A}}

\def\calC{{\mathcal C}}

\def\calS{{\mathcal S}}
\def\calT{{\mathcal T}}
\def\calU{{\mathcal U}}
\def\calV{{\mathcal V}}

\def\calX{{\mathcal X}}

\def\calZ{{\mathcal Z}}

\def\bA{{\pmb A}}
\def\bB{{\pmb B}}

\def\bH{{\pmb H}}
\def\bI{{\pmb I}}

\def\bQ{{\pmb Q}}
\def\bR{{\pmb R}}

\def\bV{{\pmb V}}
\def\bW{{\pmb W}}
\def\bX{{\pmb X}}
\def\bY{{\pmb Y}}
\def\bZ{{\pmb Z}}

\def\ba{{\pmb a}}

\def\bc{{\pmb c}}
\def\bd{{\pmb d}}
\def\be{{\pmb e}}
\def\bf{{\pmb f}}

\def\bh{{\pmb h}}

\def\bn{{\pmb n}}

\def\bp{{\pmb p}}
\def\bq{{\pmb q}}
\def\br{{\pmb r}}
\def\bs{{\pmb s}}
\def\bt{{\pmb t}}
\def\bu{{\pmb u}}
\def\bv{{\pmb v}}
\def\bw{{\pmb w}}
\def\bx{{\pmb x}}
\def\by{{\pmb y}}
\def\bz{{\pmb z}}

\newcommand{\bPhi}{  \pmb{\Phi}  }
\newcommand{\bPsi}{  \pmb{\Psi}  }
\newcommand{\bGam}{  \pmb{\Gamma}  }

\newcommand{\bTh}{  \pmb{\Theta}  }

\newcommand{\balp}{  \pmb{\alpha}  }

\newcommand{\blam}{  \pmb{\lambda}  }
\newcommand{\bmu}{  \pmb{\mu}  }

\newcommand{\bom}{  \pmb{\omega}  }
\newcommand{\bthe}{  \pmb{\theta}  }

\newcommand{\bone}{  \pmb{1}  }
\newcommand{\bzer}{  \pmb{0}  }

\newcommand{\upperRomannumeral}[1]{\uppercase\expandafter{\romannumeral#1}}

\newcommand{\st}{ \textup{s. t.} }
\newcommand{\argmin}{ \textup{argmin} }
\newcommand{\argmax}{ \textup{argmax} }

\newcommand{\rank}{ \textup{rank} }
\newcommand{\conv}{ \textup{conv} }

\def\tr{{\textup{tr}}}

\newcommand{\lrb}[1]{ \lbrace #1 \rbrace }

\newtheorem{lemma}{Lemma}

\newtheorem{proposition}{Proposition}

\newtheorem{definition}{Definition}

\newtheorem{remark}{Remark}
\newtheorem{problem}{Problem}

\begin{document}

\title{User Assignment in C-RAN Systems: Algorithms and Bounds }

\author{Hadi~Ghauch,~\IEEEmembership{Student Member,~IEEE,}
        Muhammad~Mahboob~Ur~Rahman,~\IEEEmembership{Member,~IEEE,} 
        Sahar~Imtiaz,~\IEEEmembership{Student Member,~IEEE,}
        Christer~Qvarfordt,~\IEEEmembership{Non-Member,~IEEE,}
        Mikael~Skoglund,~\IEEEmembership{Senior Member,~IEEE,} 
        and~James~Gross,~\IEEEmembership{Senior Member,~IEEE} 
		
\thanks{H. Ghauch,  S. Imtiaz, M. Skoglund and J. Gross are with the School of Electrical Engineering and the ACCESS Linnaeus Center, KTH Royal Institute of Technology, Stockholm, Sweden (E-mails: ghauch@kth.se, sahari@kth.se, james.gross@ee.kth.se, skoglund@kth.se). }
\thanks{M. Mahboob Ur Rahman was affiliated with KTH from November 2013 until April 2016. Currently, he is an Assistant Professor at the Information Technology University (ITU), Lahore, Pakistan (E-mail: mahboob.rahman@itu.edu.pk).}
\thanks{C. Qvarfordt is with Huawei, Stockholm, Sweden (E-mail: christer.qvarfordt@huawei.com). } }

\maketitle

\begin{abstract}
In this work, we investigate the problem of mitigating interference between so called antenna domains of a cloud radio access network (C-RAN). 
In contrast to previous work, we turn to an approach utilizing primarily the optimal assignment of users to central processors in a C-RAN deployment. 
We formulate this user assignment problem as an integer optimization problem, and propose an iterative algorithm for obtaining a solution. 
Motivated by the lack of optimality guarantees on such solutions, we opt to find lower bounds on the problem, and the resulting interference leakage in the network. 
We thus derive the corresponding Dantzig-Wolfe decomposition, formulate the dual problem, and show that the former offers a tighter bound than the latter. 
We highlight the fact that the bounds in question consist of linear problems with an exponential number of variables, and adapt the column generation method for solving them. 
In addition to shedding light on the tightness of the bounds in question, our numerical results show significant sum-rate gains over several comparison schemes. 
Moreover, the proposed scheme delivers similar performance as W-MMSE with a significantly lower complexity (around $10$ times less). 

\end{abstract}

\IEEEpeerreviewmaketitle

\section{Introduction}
It has  been reported that the total volume data is expected to increase by tenfold, between 2013 and 2019~\cite{Ericsson_mobility_14}, whereby mobile data constitutes the largest fraction. In view of meeting this exponentially increasing demand, 5G systems have to deliver ever-increasing data rates. Such increases are usually met by acquiring new spectrum, realizing higher spectral efficiency, and leveraging densification. 
From a historical perspective, a significant fraction of the gains in data rates are due to densification~\cite{Dohler_PhyLayer_11}, that inevitably lead more and more to the need to coordinate the interference between different base stations.

An extensive body of work suggests that in general, coordination among (clusters of) base stations is a key to achieving higher sum-rates in the network, namely, the ideas of Coordinated Multi-point (CoMP)~\cite{gesbert_comp_10, Emil:TSP:2011} and Interference Alignment (IA)~\cite{cadambe_interference_2009, maddah-ali_communication_2008}. 
However, in cellular networks of the forth generation, the overhead associated with \emph{distributed coordination algorithms} has been identified as major limiting factor to reaping of the above mentioned sum-rate gains (e.g.,~\cite{ayach_overhead_11,Lozano_limits_coop_13}). Lately, novel architectures such as \emph{Cloud Radio Access Networks (C-RAN)} have emerged as an enabler for coordinating transmit antennas in a dense deployment. The original ideas behind C-RAN, i.e., the feasibility of connected base stations (BSs), can be traced back to~\cite{Quek_CoMP_wLB_13}.
Typically, a C-RAN consists of Remote Radio-Heads (RRHs) (also referred to as \emph{radio-heads}), assumed to have limited baseband/processing capabilities, and are connected to the so-called \emph{central processors (CPs)}. Central processors act as centralized compute nodes, that gather all the required \emph{Channel State Information (CSI)} from a cluster of connected radio-heads, perform the required optimization (e.g., precoding), and send the resulting parameters to the relevant radio-heads. An \emph{Antenna Domain (AD) }is the collection of radio-heads connected to a particular central processor.~\footnote{This can also be termed as distributed antenna systems~\cite{Wang_DAS_PCSI_12}, macro-cell (in the context of HetNets)~\cite{Lee_BS_clustering_CRAN_13}, BS cluster~\cite{gesbert_comp_10}, etc.} When multiple antenna domains (multi-AD) are present in the system, this inevitably results in \emph{intra-} and \emph{inter-AD interference}, that limit the network performance. 
In this work, we focus on managing the latter, while assuming that the former is effectively handled via precoding, such that its effect can be neglected.

With the exception of~\cite{Ghauch_ADFconf_16}, all prior C-RAN work focused on managing \emph{intra-AD} interference only, by considering a network with one central processor. For instance, the authors in~\cite{Lee_BS_clustering_CRAN_13} investigated dynamic clustering of radio-heads, where users within each cluster are served in a Joint Transmission (JT)-like manner. The same model was adopted in \cite{Yu_MulticastBF_15} and \cite{Yu_MulticastBF_16},  where the authors consider the problem of forming clusters of radio-heads in the presence of caching and multi-cast transmission. A similar model for coordination was employed in~\cite{Yu_EE_CRAN_16}, focusing on energy efficient transmission instead. In~\cite{Rahman_RRHclust_15}, (looser) coordination among the radio-heads within the antenna domain was investigated, where Coordinated Beamforming (CB)-type~precoding~was employed.

Thus, managing \emph{intra-AD interference} has been addressed fairly well, in the C-RAN literature. However, the issue of \emph{inter-AD interference} in a multi-AD setup, remains open. We underline that \emph{distributed algorithms} for \emph{precoder optimization} (e.g., leakage minimization~\cite{gomadam_distributed_2011}, minimum MSE~\cite{schmidt_minimum_2009}, W-MMSE~\cite{shi_wmmse_2011}) can, in principle, be applied to mitigate inter-AD interference. 
However, state-of-the-art approaches such as W-MMSE~\cite{shi_wmmse_2011}, exhibit slow convergence (hundreds/thousands of iterations for moderate system size) and elevated complexity (per iteration).  
Moreover, increased system dimensions (e.g., users, antennas, radio-heads) result in even higher number of required iterations~\cite{Schmidt_comparison_13}, and thus increased delay. This is particularly problematic, especially when bearing in mind the small scheduling time slots envisioned in C-RAN, and lower channel coherence times (due to increased mobility). Such issues severely limit the scalability of W-MMSE-type approaches, thus making them ill-suited for densely deployed C-RANs (detailed in Sec.~\ref{sec:comm_overhead}). Finally, by alleviating the requirement that coordination algorithms be distributed, C-RAN opens-up the possibility to leverage \emph{fully centralized} approaches, to interference management: this opportunity should naturally be explored.

Our approach addresses the above limitations, by proposing a low-complexity algorithm for user assignment. This problem is (mathematically) related to the user assignment and precoder design in a MISO \emph{Interfering Broadcast Channel (IBC)}, in the context of traditional cellular networks. We thus survey the most relevant ones. Several works have addressed the problem of joint precoding and user assignment, however, for the optimization of power and energy efficiency (e.g.,~\cite{Li_minEE_15, Lu_joint_UA_prec_15}).
\cite{Hong:2013:joint} considered a HetNet setup, where each cell consisted of several inter-connected BSs, focusing however on the different problem of joint precoding and BS clustering, for sum-rate maximization.  
The authors in~\cite{Sanjabi:2014:optimal} tackle the joint optimization of user assignment and precoding, by extending the well-known W-MMSE algorithm~\cite{shi_wmmse_2011}: despite its (local) optimality guarantees, the resulting algorithm is prohibitively complex for setups smaller than the one considered here~\cite{Ghauch_DistUA_Prec_15}, and thus ill-suited even for benchmarking.

Before the popularization of the CRAN terminology, similar ideas 
 existed, under the umbrella of distributed antenna systems (DAS). In~\cite{Wang_DAS_PCSI_12,Feng_DAS_SRMax_13}, the authors consider a mixture of instantaneous and statistical CSI at the central processor. 
 Specifically,~\cite{Wang_DAS_PCSI_12} considers multiple single-antenna users in each cell
 and investigates the problem of joint beamforming design that maximizes
 the instantaneous weighted sum rate of the system. On the other hand,~\cite{Feng_DAS_SRMax_13} considers a single multi-antenna user in each cell, and solves the ergodic sum rate maximization problem in a nearly-optimal manner
 via joint optimization of (diagonal) input covariance matrices for all users. Nevertheless, both works focus on optimal/near-optimal
 precoding design, while the user assignment remains unchanged. Contrary to these works, we assume that instantaneous CSI is available at each central processor, through a fast inter-AD backhaul. Moreover, we focus on the user assignment problem, rather than precoding.   

The issue of inter-AD interference mitigation in a multi-AD C-RAN setting was first addressed in our earlier work~\cite{Ghauch_ADFconf_16}. We investigated the optimal assignment of radio-heads to antenna domains, assuming CB-type precoding within each antenna domain - that cannot fully suppress intra-AD interference. Naturally, higher performance can be achieved by assuming \emph{tighter coordination} among the radio-heads of each antenna domain, e.g., joint transmission within each antenna domain.

We consider such a setup in this work, and investigate the optimal assignment of users to antenna domains, assuming that intra-AD interference is effectively handled (by leveraging any of the aforementioned works, and thus assume it is negligible).
More specifically, we focus on theoretical aspects of the \emph{user assignment (UA)} problem: Given an initial state (i.e., assignment of users to radio-heads, and radio-heads to central processors), we study the optimal assignment of users to antenna domains, using the total interference leakage as performance metric. The main contributions of the paper are the following:
\begin{itemize}
\item We formulate the UA problem as an integer optimization problem, and then employ \emph{Block-Coordinate Descent (BCD)} to iteratively solve the problem. 
\item The lack of theoretical guarantees on the obtained solution, as well as the complicated nature of the problem, motivates us to find useful and meaningful \emph{lower bounds} on the UA problem (since it represents the total interference leakage).
For that purpose, we derive the corresponding \emph{Dantzig-Wolfe (DW)} decomposition (a Linear Program (LP) with exponentially many variables), and adapt the \emph{Column Generation Method (CGM)} to compute the DW lower bound. We shed light on the tightness of the DW decomposition, by deriving simple bounds on the error. We also derive the dual problem (a natural lower bound), and show that the DW lower bound is tighter than that of the dual problem. 
\item We provide some numerical results that highlight the performance of our proposed algorithm (we include W-MMSE in the simulation results, for benchmarking purposes only). Moreover, for typical mobility patterns of terminals, the proposed scheme is robust to some degree of outdated CSI, allowing a significant reduction of the involved overhead. 
\item The proposed algorithm provides similar performance as W-MMSE, however with a drastically lower complexity.  
\end{itemize}
Sec. II is dedicated to developing the system model, Sec. III to presenting our problem formulation and the proposed algorithm, Sec. IV to detailing the proposed relaxations/decompositions, and Sec. V to presenting/discussing numerical results.  
      
\section{System Model and Problem Statement}
\subsection{Notation}
We use bold upper-case letters to denote matrices, bold lower-case letters to denote vectors, and calligraphic letters to denote sets. Furthermore, for a given matrix $\pmb{A}$, $\sigma_{\max}[\bA]$ / $\sigma_{\min}[\bA]$ denote the largest/smallest singular value, $\tr(\bA)$ denotes its trace, $[\pmb{A}]_{i:j}$ denotes the matrix formed by taking columns $i$ to $j$ of $\pmb{A}$,  $\Vert \pmb{A} \Vert_F^2$ its Frobenius norm,  $\bA^T$ its transpose, and $\pmb{A}^\dagger$ its conjugate transpose. $ [\bA]_{i,j} = a_{i,j}$ denotes element $(i,j)$ of matrix $\bA$, and $[\ba]_i$ element $i$ of vector $\ba$. For any two vectors $\bx, \by$ (resp. matrices $\bX, \bY$), inequalities such as $\bx \leq \by$ (resp. $\bX \leq \bY$) hold element-wise.   
While $\bI_n$ denotes the $n \times n$ identity matrix, 
  $\pmb{1}$ denotes the all-one vector, 
  $\pmb{0}$ denotes the all-zero vector, of appropriate dimension. 
Moreover, $\be_n$ is the $n$th vector of the canonical basis (having appropriate dimension), 
  $\IB$ denotes the binary set, and
  $\IZ_{+}$ denotes the set of natural numbers.
Given a set $\calX$, $|\calX|$ denotes its cardinality, and $\conv(\calX)$ its convex hull. 
Given an optimization problem $(P)$, series of equivalent ones are denoted as $(P_2)$, $(P_3)$, etc.

\subsection{Model and Assumptions} 
\label{sec:model_assumptions}
We consider the operation of a cloud radio access network (C-RAN) deployment.
Assume a large area, comprising of $A$ \emph{central processors}, $N_{\textup{T}}$ \emph{remote radio-heads}, and $U_{\textup{T}}$ \emph{users}. 
Each central processor is connected to $N$ radio-heads via wireless/wired links, where each radio-head is serving a set of users. 
We refer to the collection of radio-heads connected to each central processor as an \emph{antenna domain.} 
Thus, each antenna domain is serving a set of users (thereby abstracting the operation of the radio-heads in the system). 
Each antenna domain comprises of $N$ radio-heads and $U$ users.
\footnote{While quantities such as $U$, $N$, $\calU$ vary across different antenna domains (i.e., $U_j$, $N_j$, $\calU_j$), we drop the subscript for notation simplicity (without loss of generality). We also drop any time-related indexes.}
We denote by $\calA$ the set of central processors, $\calU$ the set of users served by antenna domain $j \in \calA$, and  $\calU_{\textup{T}}$ the set of all users, i.e., $\calU_{\textup{T}} = \lrb{ j_n \ | \ j \in \calA , \ n \in \calU } $. 
We assume that each radio-head is equipped with $M$ antennas, while users have a single antenna each.

\begin{figure}
  \begin{center}
  \begin{minipage}{7cm}
    \centering
    \includegraphics[scale=.45]{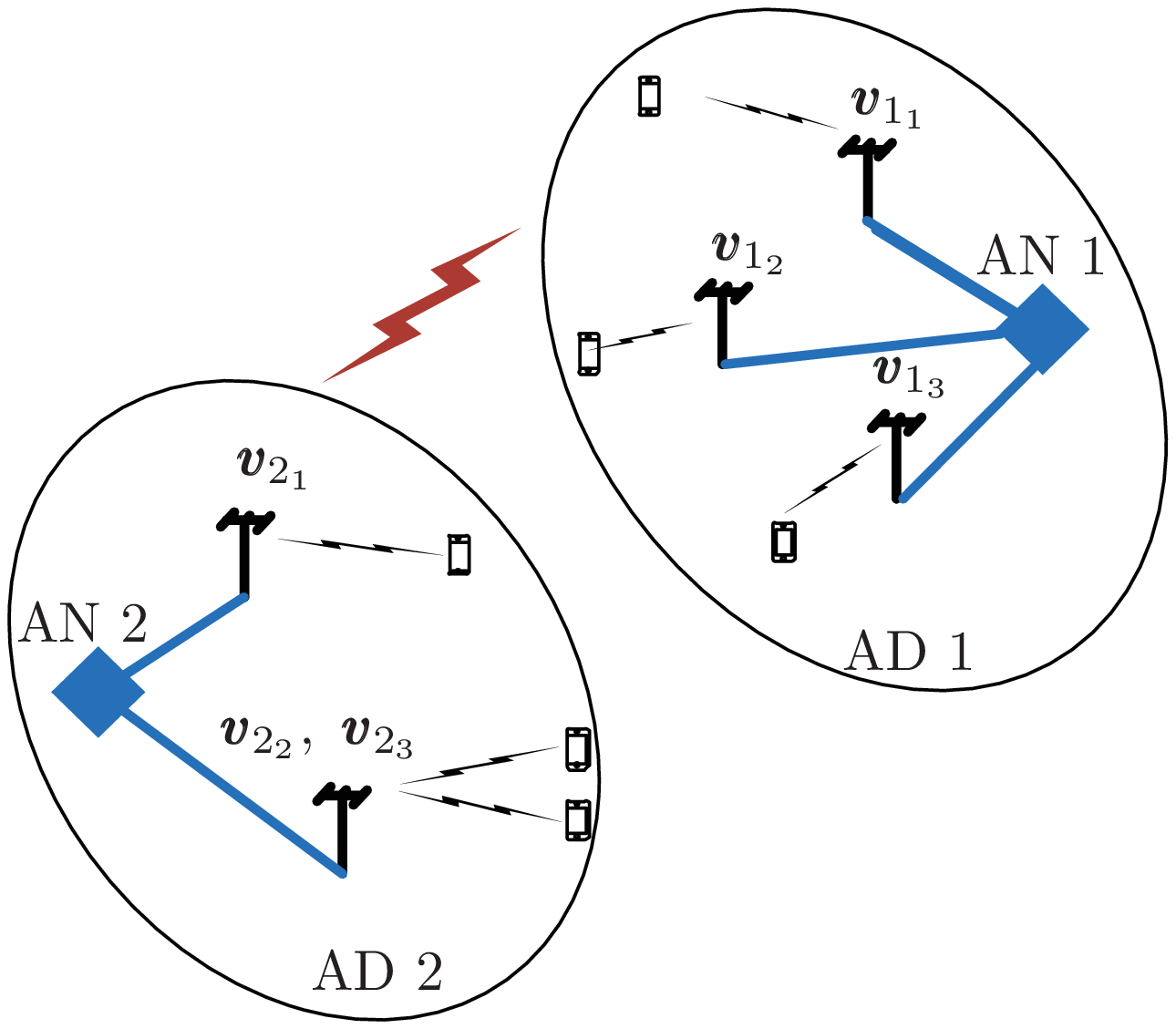} 
    \caption{System Model: $\bv_{1_1}, ..., \bv_{1_3} $ are the precoders used by AD 1 to serve its users, and $\bv_{2_1}, ..., \bv_{2_3} $ are the precoders used by AD 2 to serve its users  }%
    \label{fig:sysmod}
  \end{minipage}
  \hspace{1cm}
  \begin{minipage}{7cm}
    \centering
    \includegraphics[scale=.55]{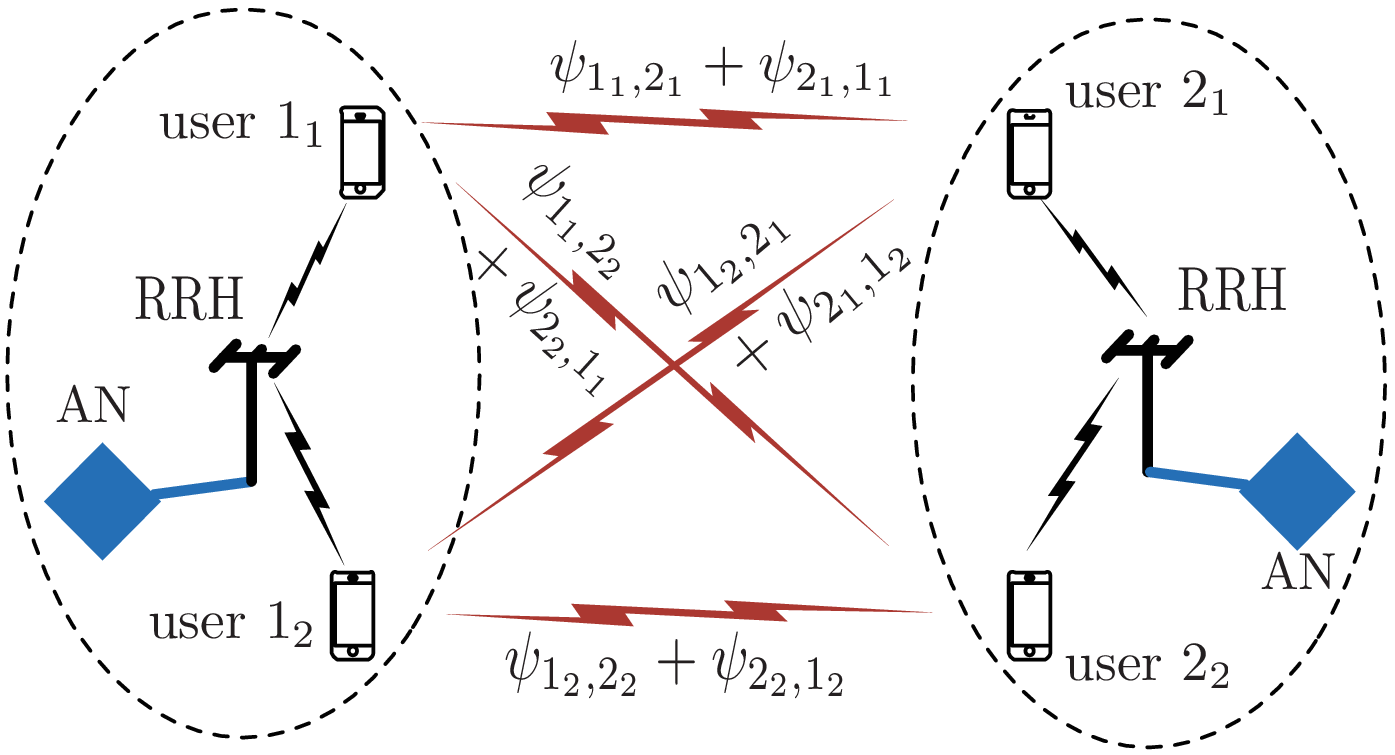}    
  \caption{Toy Example (interference marked in red): total interference equal to inter-AD interference }
    \label{fig:toyex}
  \end{minipage}
  \end{center}
\end{figure}

Central processors in C-RAN perform generally all digital processing in uplink and downlink. 
Note that in the following we only consider downlink operations. 
Thus, they gather CSI from all the users (via the radio-heads), perform the required scheduling/optimization of radio resources, and communicate the chosen resource allocations to the radio-heads along with the payload data to be transmitted. We further assume that central processors can communicate with each other over a fast backbone, without quantifying the cost of this communication in the considered optimization problem. This enables the availability of global and perfect CSI at each of the central processors, at the beginning of each scheduling time-slot. 
We assume full-buffer mode for all users, i.e. there is a continuous backlog of data for each user in the system to be transmitted by the cellular network.
The different radio-heads within each antenna domain are assumed to be tightly synchronized, i.e., at the carrier level, essentially acting as a large virtual antenna array.
A main motivation for C-RAN is to find ways of leveraging global CSI, in order to manage antennas in a dense deployment (thus requiring such an assumption). We also note that any algorithm that coordinates antennas across different central processors, is likely to require such levels of CSI.   
We underline at this point that such assumptions are widespread in C-RAN related performance studies (e.g., \cite{Yu_MulticastBF_15,Yu_MulticastBF_16,Lee_BS_clustering_CRAN_13}).  
Moreover, their implications are further discussed in Sec.~\ref{sec:comm_overhead}. 
We further assume that precoding within each antenna domain is designed to reduce intra-AD interference. For instance, this may be accomplished by finding the zero-forcing or leakage minimizing precoder~\cite{gomadam_distributed_2011}, for intra-AD users. We thus assume that interference is negligible for users within the same antenna domain. We briefly discuss means of realizing this design in the numerical results, and Appendix~\ref{prop:opt_prec}. Fig.~\ref{fig:sysmod} shows the resulting system model.

We assume that an initial assignment of users to antenna domains already exists. 
Hence, $j_n$ denotes the index of the $n$th user, in the $j$th antenna domain, $j_n \in \calU_{\textup{T}}$. 
Then, its received signal is given by
\begin{align} \label{eq:sig_mod}
y_{j_n} =    \sum_{q \in \calU } \bh_{j,j_n} \bv_{j_q} s_{j_q} \sqrt{p_j} + \sum_{ \substack{ i_m \in \calU_{\textup{T}} \\ i \neq j} }  \bh_{i,j_n} \bv_{i_m} s_{i_m} \sqrt{p_i} + n_{j_n}  
\end{align}
where $\bh_{i,j_n} \in \IC^{1 \times MN}$ is the (MISO) channel from antenna domain $i$ to user $j_n$, $\bv_{i_m} \in \IC^{MN \times 1}$ the beamforming vector to user $i_m \in \calU_{\textup{T}} $, $s_{i_m}$ the data symbol for user $i_m \in \calU_{\textup{T}}$ such that $\IE[s_{i_m} s_{i_m}^\dagger ] = 1 $,  $n_{j_n}$ the AWGN noise for user $j_n \in \calU_{\textup{T}}$ such that $\IE[n_{j_n} n_{j_n}^\dagger ] = \sigma_{j_n}^2 $,  and $p_j$ the transmit power for antenna domain $j \in \calA$ (where we assume equal power allocation among users of the same antenna domain). While the first term in~\eqref{eq:sig_mod} represents intra-AD interference, the second one denotes inter-AD interference.   

Moreover, neglecting the intra-AD interference, we rewrite the received signal and SINR as, 
\begin{align}
y_{j_n} &=   s_{j_n} \sqrt{p}_j + \sum_{ \substack{ i_m \in \calU_{\textup{T}} \\ i_m \neq j_n } }  \bh_{i,j_n} \bv_{i_m} s_{i_m} \sqrt{p}_i + n_{j_n} \nonumber \\
\gamma_{j_n} &=  \frac{  p_j }{ \sum_{ \substack{ i_m \in \calU_{\textup{T}} \\ i_m \neq j_n } }  p_i |\bh_{i,j_n} \bv_{i_m} |^2 + \sigma_{j_n}^2 } \quad . \label{eq:sinr}
\end{align}
We define the SNR of user $j_n \in \calU_{\textup{T}} $ as  $ \textrm{SNR}_{j_n} =  p_j /\sigma_{j_n}^2 $ . 
Note that the above signal model, SINR, and sum-rate are identical to those of a MISO IBC. 
In that sense, `antenna domain' and `cell' are mathematically equivalent. However, in the C-RAN literature the nomenclature for `cell' and `antenna domain' are purposely distinguished: Indeed, one of the main motivations for C-RAN is to resolve the traditional `cell' concept.

\subsection{Problem Statement} 
\label{sec:problem_statement}
Assuming optimal encoding/decoding, and treating interference as noise, the achievable sum-rate of the network is given by
\begin{align} \label{eq:sum_rate}
R_{\Sigma} =\sum_{j_n \in \calU_{\textup{T}} } \log_2(1 + \gamma_{j_n} ) \quad .
\end{align}

In this work we are interested in maximizing the sum-rate of the network. 
Note that in general, it is well known that the multi-cell multi-user sum-rate optimization problem is NP-hard~\cite{Razaviyayn_dof_2011}. 
Thus, tackling `surrogate problems' to sum-rate maximization is an inevitable next step: 
This is evidenced by the overwhelming number of previous works, in the context of \emph{multi-cell coordination}, focusing on minimizing the interference leakage\cite{gomadam_distributed_2011}, minimizing the total MSE~\cite{schmidt_minimum_2009}, maximizing per-stream SINR~\cite{gomadam_distributed_2011}, and only a few works that have directly addressed the sum-rate problem (e.g.,~\cite{santamaria_maximum_2010,shi_wmmse_2011}). Moreover, approaches that attempt to maximize the sum-rate (e.g., W-MMSE~\cite{shi_wmmse_2011} and W-MMSE with UA~\cite{Sanjabi:2014:optimal}), suffer from slow convergence and elevated complexity that make them unfit for ultra-dense networks - which are quite prevalent in future cellular networks~\cite{METISD62}.

With the above in mind, we optimize the \emph{interference leakage}.  
We focus on sub-optimal but fast heuristics (with some analytical guarantees), especially since the amount of remote radio-heads, users and antenna domains is likely to be quite high in C-RAN systems. It will also become clear that despite this simplification, the resulting optimization problems are still quite challenging.


\section{Proposed Approach} 
We assume that initially, each antenna domain selects precoders that reduce the intra-AD interference.  
Our approach focuses on the user assignment step that is chosen jointly among the antenna domains, to further minimize the interference leakage.

\subsection{Interference Coupling Coefficients and User Assignment} 
\label{sec:motivation}
We denote by $\psi_{i_m, j_n }$ the so-called \emph{interference coupling coefficient} between users $i_m$ and $j_n$,  
\begin{align*}
  \psi_{i_m, j_n } =
   p_i| \bh_{i,j_n} \bv_{i_m}  |^2 , \ \forall \ (i_m, j_n) \in \calU_{\textup{T}}^2 , \ \ i_m \neq j_n  \quad . 
\end{align*}
$\psi_{i_m, j_n }$ denotes the interference that the transmission to user $i_m \in \calU_{\textup{T}} $, causes to user $j_n \in \calU_{\textup{T}} $ (recall that $ \psi_{i_m, j_n } \neq  \psi_{j_n , i_m}  $). 
Let $\bPsi \in \IR_+^{U_{\textup{T}} \times U_{\textup{T}}} $ be the matrix formed by gathering all the coupling coefficients 
\begin{align}
  [\bPsi]_{j_n, i_m} =
  \begin{cases}
        \psi_{j_n, i_m} , \ \forall i \neq j \\
        0 , \forall i = j    
  \end{cases},
  \forall \ (j_n,i_m) \in \calU_{\textup{T}}^2 \quad .
\end{align}
With that in mind, $g_{j_n}$, the total interference leakage seen by user $j_n \in \calU_{\textup{T}}$, is given by,
\begin{align} 
g_{j_n}   &\triangleq \sum_{k \in \calA } \sum_{ \substack{ l \in \calA \\ l \neq k } } \left(  \sum_{ \substack{ i_m \in \calU_{\textup{T}} \\ i_m \neq j_n } }  x_{k,i_m}  \psi_{i_m, j_n} x_{l,j_n}   \right)
\end{align}
where $ x_{k, j_n } \in \lrb{0, 1} ,  \ \forall \  j_n \in \calU_{\textup{T}} , \ \ k \in \calA$
be the assignment variable for user $j_n$ to antenna domain $k$. 
Recall that we previously assumed that user $j_n$ is already assigned to antenna domain $j$. 
Hence, $x_{k, j_n } $ models the potential re-assignment of user $j_n $ (assigned to antenna domain $j$ initially), to antenna domain $k$. 
Various initial assignments could be utilized. 
For instance, this could relate to the assignment from the previous period. 
Alternatively, an assignment based on the strongest channels could be used to initialize the system.
Note in the above equation that we only account for interference, when users $i_m$ and $j_n$ are in different antenna domains (since intra-AD interference is assumed negligible, through an appropriate choice of precoding).
The total interference leakage, $f$, is then defined as $f \triangleq   \sum_{j_n \in \calU_{\textup{T}}}  g_{j_n}   $, and re-written as,  
\begin{align}\label{eq:lkg_joint}
 f \triangleq \sum_{k \in \calA } \sum_{ \substack{ l \in \calA \\ l \neq k }  }  \left( \sum_{ j_n \in \calU_{\textup{T}} } \sum_{ \substack{ i_m \in \calU_{\textup{T}} \\ i_m \neq j_n }} x_{k,i_m}  (p_i \bh_{i, j_n} \bv_{i_m} \bv_{i_m}^\dagger \bh_{i, j_n}^\dagger  ) x_{l,j_n} \right)  \quad .
\end{align}
Note that a \emph{joint optimization} of the set of assignment variables $\lrb{ x_{k,i_m} }$, and the set of precoding vectors $\lrb{ \bv_{i_m} }$, is infeasible. 
Moreover, an \emph{alternating minimization} approach is of little interest, as explained below (Remark~\ref{rmk:alt_opt}). 
\begin{remark}\label{rmk:alt_opt} \rm
Looking at~\eqref{eq:lkg_joint} intuitively suggests an alternating optimization, where $\lrb{ x_{k,i_m} }$ and $\lrb{ \bv_{i_m} }$ are alternately and iteratively optimized. 
However, convergence for such an algorithm \emph{cannot} be proven. 
It is well-understood that convergence of alternating optimization can only be shown when both the precoding and assignment subproblems are solved to optimality~\cite{Razaviyayn_BCD_12}.
Clearly, it is not possible to solve the latter using any polynomial-time algorithm, as it is NP-hard. 
This limitation from alternately optimizing both quantities, is the chief reason for focusing on the assignment problem only in the following. Additionally, while an alternating optimization approach would yield better performance, it would naturally incur larger complexity and communication overhead. We recall that this goes against our main design criteria, of suboptimal yet fast solutions to leakage minimization.  
\end{remark} 
 
Our focus is thus on the assignment problem. Despite this apparent simplification, a main finding of this work is that the problem is still intractable. 
Considering the user assignment only, we thus express the total leakage as a function of all the assignment variables, $ \lrb{x_{k,i_m}} $:
\begin{align} \label{eq:f_lkg}
f( \lrb{ x_{k,i_m} } ) \triangleq \sum_{k \in \calA } \sum_{ \substack{ l \in \calA \\ l \neq k }  }  \left( \sum_{ j_n \in \calU_{\textup{T}} } \sum_{ \substack{ i_m \in \calU_{\textup{T}} \\ i_m \neq j_n }} x_{k,i_m}  \psi_{i_m, j_n} x_{l,j_n} \right)  \quad .
\end{align} 
Recall that due to the assumed precoding, the inter-AD interference leakage coincides with the total interference leakage, $f$, in the system. We have illustrated all the above in a simple toy example (Fig.~\ref{fig:toyex}). 


\begin{problem}[User Assignment (UA)] \label{def:UAprob} \rm
Given an initial state (i.e., assignment of users to radio-heads, and radio-heads to central processors), the UA problem is given by the optimal assignment of users to antenna domains, w.r.t. minimizing the total interference leakage in the system. The corresponding optimization problem results in the following integer program:
\begin{align} 
(P)\begin{cases} 
   \underset{\lrb{x_{k,j_n}  }}{\min} \ f = \underset{ k \in \calA }{\sum} \ \underset{ l \neq k }{\sum} ( \underset{ j_n \in \calU_{\textup{T}} }{\sum} \ \underset{ i_m \neq j_n }{\sum} x_{k,i_m}  \psi_{i_m, j_n} x_{l,j_n}  )    \nonumber \\
   \st \ \sum_{i_m \in \calU_{\textup{T}} }  x_{k,i_m} = \rho_k , \ \forall k \in \calA ~,~~ \sum_{k \in \calA } x_{k, i_m} \leq 1 , \ \forall i_m \in \calU_{\textup{T}}  \nonumber \\
   \hspace{.7cm}  x_{k,i_m} \in \lrb{0,1} , \ \forall (k, i_m) \in \calA \times \calU_{\textup{T}} \nonumber \quad .
\end{cases}
\end{align}
\end{problem}
The first constraint specifies that $\rho_k \in \IZ_{+}$  users are to be assigned to each antenna domain (where $\rho_k < U$), i.e., the \emph{loading constraint}. 
We introduce this constraint for the sake of load balancing (i.e., to prevent highly asymmetric cases where all users get assigned to one antenna domain, while the rest are idle).
Moreover, the second constraint,  i.e., the \emph{assignment constraint}, ensures that each user is assigned to \emph{at most} one antenna domain. 

We first start by rewriting $(P)$ in vector and matrix form - both of which will be used later. 
Let $\bx_k = [ x_{k,1_1} , \cdots , x_{k,A_U} ]^T , ~ \ \bx_{k} \in \IB^{U_{\textup{T}}} ,  \forall k \in \calA $ denote the \emph{assignment vector} for central processor $k$, and $\bX$ the \emph{aggregate assignment matrix} for the system, i.e., $\bX = [ \bx_1 , ...., \bx_A ] \in \mathbb{B}^{U_{\textup{T}} \times A }$. 

\begin{proposition} \label{prop:prob_form}
$(P)$ can be rewritten in equivalent vector form, 
\begin{align} 
(P_2)\begin{cases} 
   \min f(\lrb{ \bx_k } ) = \sum_{ k \in \calA } \sum_{l \neq k}  \bx_k^T \bPsi \bx_l   \nonumber \\
   \st \  \  \sum_{k \in \calA } \bx_{k} \leq \bone , \ \pmb{1}^T  \bx_k = \rho_k , \  \bx_k \in  \mathbb{B}^{U_{\textup{T}}}, \  \ \forall k \in \calA  \nonumber \\   
\end{cases}
\end{align}
and matrix form 
\begin{align} 
(P_3)\begin{cases} 
   \min f(\bX ) = \tr ( \bX^T \bPsi \bX \pmb{\Omega}  ) \ \ \st \ \bX \pmb{1} \leq\pmb{1} , \ \ \bX \in  \calS_{\rho}  \quad . \nonumber
\end{cases}
\end{align}
$\calS_{\rho}$ denotes the set of all $U_{\textup{T}} \times A$ binary matrices, that satisfy the loading constraint, 
$\calS_{\rho} \triangleq \lrb{ \bQ \in \IB^{U_{\textup{T}} \times A } \ |  \ \bQ^T \pmb{1}= \pmb{\rho}  }$
where $\pmb{\rho} \triangleq [\rho_1, ..., \rho_A ]^T  $, and $\pmb{\Omega} \triangleq \bone \bone^T - \bI_A $.
\end{proposition}
\begin{IEEEproof} 
The derivations are shown in Appendix~\ref{sec:app_prob_form}. 
\end{IEEEproof}
It can be seen from $(P_2)$ that $f$ is not jointly convex in all the variables, due to the coupling among them. 
However, we underline the inherent \emph{multi-linear} nature of $f$ (taken separately in each variable, $f$ is linear), that we exploit for the optimization.  
We underline that the UA problem and solution take a simple form, in the case of two antenna domains. We have thus proposed an equivalent of $(P_2)$ that enables a straightforward and systematic solution (all details are relegated to Appendix~\ref{sec:special_cases}). 

\subsection{Proposed Algorithm}
We first present the following definition.
\begin{definition}[Integrality Property for LP] \rm \label{def:integrality} 
Consider the binary linear program: 
$$ (BLP) \ \bx^\star = \min_{\bx} \ \bc^T \bx , \ \st \ \bx \in \calC , \ \ \bx \in \IB^N, $$
and its \emph{continuous relaxation} (CR) (also known as LP relaxation), 
$$ (CR) \ \hat{\bx} = \min_{\bx} \ \bc^T \bx , \ \st \ \bx \in \calC , \ \ \bzer \leq \bx \leq \bone , $$
where $\calC$ is a convex set. $\calC$ is said to satisfy the \emph{integrality property} if all its vertexes correspond to binary vectors: Then, the so-called \emph{continuous relaxation} (CR) is optimal \cite{Frangioni_Lagrangian_05}, i.e., $\hat{\bx} = \bx^\star$. 
\end{definition}

Due to the coupled nature of the objective function of $(P_2)$, we leverage the well known Block-Coordinate Descent (BCD) method, that has been applied to several areas of signal processing, e.g., transceiver optimization in cellular networks~\cite{shi_wmmse_2011,schmidt_minimum_2009,Ghauch_IWU_15,gomadam_distributed_2011}. 
In what follows, $n$ denotes the iteration number, i.e., $\bx_k^{(n)}$ denotes the value of $\bx_k$ at the $n$th iteration. 
We denote by $\bz_k^{(n)} = \lrb{ \bx_{1}^{(n+1)},..., \bx_{k-1}^{(n+1)}, \bx_{k+1}^{(n)} ,..., \bx_{A}^{(n)} }$ the block of fixed variables, for the update of antenna domain $k$, at the $n$th iteration. 

We let $f( \bx_k, \bz_k^{(n)} )$ denote the function $f(\bx_k)$, when the variables in block $\bz_k^{(n)}$ are fixed, which can be written as 
\begin{align} \label{eq:rk}
f( \bx_k, \bz_k^{(n)} ) &= \bx_k^T \bPsi  \left(  \sum_{l = 1}^{k-1}  \bx_l^{(n+1)} + \sum_{l = k+1}^A \bx_l^{(n)} \right) \triangleq \bx_k^T \br_k^{(n)} 
\end{align}
where $\br_k^{(n)}$ is referred to as the \emph{residual} of antenna domain $k$, at the $n$th iteration. Looking at the above equation, 
 $f( \bx_k, \bz_k^{(n)} )$  is linear in $\bx_k$, implying that $f$ is linear in each block of variables. 
The application of BCD yields the following update for $\bx_k$, at the $n$th iteration. 
\begin{align} 
&\bx_k^{(n+1)} = \begin{cases} \label{opt:xk_opt0} 
   \underset{ \bx_k }{\argmin}  \ f(\bx_k,  \bz_k^{(n)} ) ~~~~\st \ \pmb{1}^T \bx_k = \rho_k , \ \bx_{k} \leq \bom_k ,  \ \bx_k \in  \mathbb{B}^{U_{\textup{T}}}  
\end{cases} 
\end{align}
where $\bom_k \triangleq \bone - \sum_{l \neq k} \bx_l $ is the set of feasible assignments for $\bx_k$.  
The above problem belongs to the class of Mixed-Integer Linear Programs (MILPs). Moreover, it is a special case of the \emph{generalized assignment problem (GAP)}. Though the generic formulation of GAP is known to be NP-hard, we exploit the particular structure of~\eqref{opt:xk_opt0} (i.e., the integer constraints are binary ones), to show that it is equivalent to a LP. Let $\calC$ be the set formed by the first two constraints in~\eqref{opt:xk_opt0}, i.e., $\calC = \lrb{ \pmb{1}^T \bx_k = \rho_k , \ \bx_{k} \leq \bom_k  } $. 
Note that $\rho_k$ is integer (by definition) and $\bom_k$ is a binary vector (by construction). 
Then, a simple argument can be put forth to show that the vertexes/extreme points of $\calC$ can only be binary vectors (the proof is similar to~\cite{Frangioni_Lagrangian_05}), and thus satisfies the integrality property: its continuous relaxation yields the optimal solution (detailed in Definition~\ref{def:integrality}). Another way to show that~\eqref{opt:xk_opt0} satisfies the integrality property is to write its feasible set in the form, $ \bA \bx_k \geq \bc $, and verify that $\bA$ is a totally unimodular matrix. Thus, the~last~problem~is~equivalent~to,
\begin{align} 
\bx_k^{(n+1)} = \begin{cases} \label{opt:xk_opt} 
   \underset{ \bx_k }{\argmin}  \ f(\bx_k,  \bz_k^{(n)} )  ~~~~\st \ \pmb{1}^T \bx_k = \rho_k , \ \bzer \leq \bx_{k} \leq \bom_k   
\end{cases}
\end{align}

As seen from~\eqref{eq:rk}, when $ \lrb{\bx_l}_{l\neq k}$ are fixed, the cost function decouples in $\bx_k$. 
Thus, given global knowledge of $\bPsi$ and assignments from all other central processors, then the optimal update for $\bx_k$ is done \emph{locally} at antenna domain $k$ (by solving~\eqref{opt:xk_opt}, a linear program). 
The optimal update for $\bx_k$ at antenna domain $k$, is a function of the assignments at all the other antenna domains (that thus have to be shared): Given assignments from other antenna domains, $(\bx_{1}^{(n+1)},..., \bx_{k-1}^{(n+1)}, \bx_{k+1}^{(n)} ,..., \bx_{A}^{(n)} )$, antenna domain $k$ forms the residual $\br_k^{(n)}$, and can proceed to solve the corresponding LP, and update $\bx_k^{(n+1)}$. The process is formalized in Algorithm~\ref{alg:1}. 
Note that the proposed approach could also be realized centrally at one central processor, if the corresponding channel knowledge is provided to that point of computations.
In the following we nevertheless present the user assignment algorithm as a decentralized approach.
\begin{algorithm} [h]
\caption{UA via BCD} \label{alg:1}
\begin{algorithmic}
\State \textbf{Input:} $ \bPsi, \ U_{\textup{T}} , \pmb{\rho} , \ A $  
\For{$n=0,1, \cdots , L_{UA}-1$}
\State // \emph{procedure at each central processor }
\State obtain   $(\bx_{1}^{(n+1)},..., \bx_{k-1}^{(n+1)}, \bx_{k+1}^{(n)} ,..., \bx_{A}^{(n)} )$  at antenna domain $k$
\State compute residual $\br_k^{(n)}$ using~\eqref{eq:rk}
\State compute feasible assignment $\bom_k$ using~\eqref{opt:xk_opt0} 
\State compute $\bx_k^{(n+1)}$ as solution to~\eqref{opt:xk_opt}
\EndFor 
\State \textbf{Output:} $ \bX^{(L_{UA})} = [ \bx_1^{(L_{UA})} , ..., \bx_A^{(L_{UA})} ] $  
\end{algorithmic}
\end{algorithm} 
\vspace{-2em}

\subsection{Convergence}  \label{sec:conv}
Let $\lrb{ \bx_k^{(n)} }_n $  be the sequence of iterates produced by the BCD in~\eqref{opt:xk_opt}, and $ \bx_k^o  \triangleq \lim_{n \rightarrow \infty}  \lrb{ \bx_k^{(n)} }_n , $ $ \forall  ~k\in \calA$. The monotonic nature of the BCD iterates was established in our earlier work~\cite{Ghauch_ADFconf_16}, and is presented below for completeness.  
\begin{lemma}[Monotonicity] \label{lem:mono}
 With each update $ \bx_k^{(n)}  \rightarrow \bx_{k+1}^{(n)} $, $f$ is non-increasing. Moreover, the sequence of function iterates $ \lrb{ f(\bx_1^{(n)}, ..., \bx_A^{(n)} ) }_n $ converges to a limit point $f(\bx_1^o,  \cdots , \bx_A^o  ) $.    
\end{lemma}
\begin{IEEEproof} Refer to Appendix~\ref{sec:app_mono}
\end{IEEEproof}
Although the above result establishes the convergence of the proposed BCD method, it only establishes convergence to a limit. However, showing that this limit is a stationary point of $f$ is not possible under the BCD framework, due to the coupled nature of the assignment constraint in $(P_2)$. Even the strongest BCD convergence results such as~\cite{Tseng_convBCD_01} cannot establish that. Therefore, we resort to finding lower bounds on the leakage problem, which we develop in the next section. But before that, we outline the system-level operation of Algorithm~\ref{alg:1}. 

\subsection{System-Level Operation of Proposed Algorithm} \label{sec:syslev_oper}

Starting from a given deployment of central processors, radio-heads and users, users are initially assigned to antenna domains, for example based on strongest channels.
The method described in this section, is merely a `wrapper' that describes how Algorithm~\ref{alg:1} operates, form a system-level perspective, shown below. 
\begin{mdframed} 
\textbf{System-Level Operation of Proposed Approach} \newline
\noindent \emph{Start with a given users-to-antenna domain assignment}  \newline  
\noindent \emph{Start by computing precoders at each central processor } \newline 
For each scheduling time-slot: 
\begin{itemize}
\item[1.] Compute $\bPsi$ at each central processor (based on CSI and precoders)  
\item[2.] Compute UA solution, $\bX^{(L_{UA})}$, at each central processor (using Algorithm~\ref{alg:1})
\item[3.] Assign users to antenna domains, based on UA solution 
\item[4.] Match preselected precoders to updated user assignment 
 \end{itemize} 
\end{mdframed}
Note that the precoders are initializers for Algorithm~\ref{alg:1}, and not updated iteratively within it. 



\section{Relaxations and Performance Bounds}
Clearly, problems such as $(P)$ are quite challenging: Despite the widespread effectiveness of methods such as BCD, even local optimality of the solution cannot be established. 
Thus, if we are given a solution using Algorithm~\ref{alg:1}, we cannot know how `good' or `bad' it might be. 
To compensate for such shortcomings, finding meaningful lower bounds on $(P)$ is of interest. 
In the following, for the problem at hand we derive the corresponding Dantzig-Wolfe (DW) decomposition, and establish that although the resulting problem is a LP, it has exponentially many variables. 
We thus adapt the Column Generation Method (CGM), for our particular problem. 
We also derive the dual problem for $(P)$, and show that it yields a looser lower bound on $(P)$.
In the following, we rely on some definition and notations:
\begin{definition}[Special LPs] \label{sec:app_specLP} \rm
Consider the following LP, 
\begin{align*}	
	(LP) \ \bx^\star = \underset{\bx \in \IR^{n} }{\argmin} \ \bc^T \bx  , \ \st \   \bone^T \bx =1 , \ \bx \geq \bzer  \quad.
\end{align*}
Let $\calV$ be the set of vertexes for $(LP)$. Then, $\calV = \lrb{\be_i }_{i=1}^n $, where $\be_i$ is the $i$th elementary vector in $\IR^{n}$. Moreover, for LPs, the optimal solution lies within $\calV$ - a fundamental result for LPs. 
\begin{align*}	
	(LP) \ \bx^\star  &=  \argmin \ \bc^T \bx  , \ \st \ \bx \in \calV \ \  \Leftrightarrow \ i^\star = \ \argmin_{ 1 \leq i \leq n } \ \bc^T \be_i  
\end{align*}
and consequently, $\bx^\star = \be_{i^\star} $.
Therefore, the solution reduces to searching over the cost $\bc$. 

\end{definition}

We also define the following notation: 
\begin{align}
  \calS_{\rho} &\triangleq \lrb{\bQ_j}_{j=1}^S , \ S = |\calS_\rho | , \ \   \alpha_j \triangleq  \tr ( \bQ_j^T \bPsi \bQ_j \pmb{\Omega}  ) , \ \forall  j = 1, \cdots, S  \nonumber \\
  \bq_j &\triangleq \bQ_j \pmb{1}  - \bone , \ \forall \bQ_j \in \calS_\rho ,  \ \ \bq_j \in \IZ^{U_{\textup{T}}} \nonumber  \\ 
  \bGam &\triangleq [\bq_1, \cdots , \bq_S ] , \ \bGam \in \IZ_{+}^{U_{\textup{T}} \times S } \label{eq:definitions}
\end{align}


\subsection{Dantzig-Wolfe Decomposition} \label{sec:DWD}
Initially proposed in their seminal paper~\cite{Dantzig_Decompostion_60}, the Dantzig-Wolfe decomposition has been widely adopted by the operations research community for finding bounds on integer programming problems. 
Based on our above definitions in~\eqref{eq:definitions}, we can rewrite $\calS_{\rho}$ and $(P_3)$ as,
\begin{align} \label{eq:DWD_binary} 
\calS_{\rho} = \lrb{ \bX = \sum_{j=1}^S w_j \bQ_j \ | \  \sum_{j=1}^S w_j = 1, \ w_j \in \IB , \forall j  }
\end{align}  
\begin{align} 
(P_4)\begin{cases} 
   \min \ f(\bX ) = \tr ( \bX^T \bPsi \bX \pmb{\Omega}  )  \ \ \  \st \ \ \bX \in \calS_{\rho}  , \ \ \bX \pmb{1} \leq \pmb{1} \nonumber
\end{cases}
\end{align} 

The above problem is still difficult to tackle, due to the combinatorial nature of $\bX \in \calS_{\rho} $ . 
The DW decomposition proceeds by relaxing $\bX \in \calS_{\rho} $ into a convex one, by taking its \emph{convex hull},  
\begin{align} \label{eq:mapping_DWD} 
\conv(\calS_{\rho}) = \lrb{ \bX = \sum_{j=1}^S w_j \bQ_j \ | \ \bone^T \bw = 1 ,  \ \bzer \leq \bw }  \quad.
\end{align}  
As a result, every point in $\conv(\calS_{\rho})$ is represented as a \emph{convex combination} of the \emph{extreme points} of $\conv(\calS_{\rho})$. Since $\calS_{\rho} \subseteq \conv(\calS_{\rho}) $, the DW problem is a lower bound on $(P_4)$, 
\begin{align} \label{opt:mat_adf_1}
(P_{\textup{DW}})\begin{cases} 
   \min \ f(\bX ) = \tr ( \bX^T \bPsi \bX \pmb{\Omega}  )  \ \ \ \st \ \ \bX \in \conv(\calS_{\rho} ) , \ \ \bX \pmb{1} \leq \pmb{1}   \quad.
\end{cases}
\end{align} 
Note that the assignment constraint can be written in an equivalent form,
\begin{align*}
\bX \pmb{1} \leq \pmb{1} &\Leftrightarrow (\sum_j w_j \bQ_j) \pmb{1} \leq \pmb{1} 
 \Leftrightarrow \sum_j w_j \bq_j + (\sum_j w_j) \pmb{1}  \leq \pmb{1} \Leftrightarrow \bGam \bw   \leq \pmb{0}_{U_{\textup{T}}}
\end{align*}
where the last one follows from the fact that $ \sum_j w_j = 1 $ (as defined by the DW decomposition). 
Moreover, recalling that $\alpha_j \triangleq  \tr ( \bQ_j^T \bPsi \bQ_j \pmb{\Omega}  ) , \ \forall  j $, and letting $\bw = (w_1, \cdots , w_S )^T$,~\eqref{opt:mat_adf_1} becomes,
\begin{align} \label{opt:mat_adf_3}
(P_{\textup{DW}})\begin{cases} 
   \underset{\bw}{\min} \ f_{\textup{DW}}(\bw) = \balp^T \bw   \nonumber \\
   \st \ \bGam \bw \leq \bzer , \ \ \bone^T \bw = 1, \ \ \bw \geq \bzer   \quad.    \nonumber
\end{cases}
\end{align}
A few remarks are in order at this stage. 
Note that despite the combinatorial and non-convex nature of $(P)$, the DW always results in a linear program (provided that $\calS_{\rho}$ is a bounded polyhedron). 
However, there is the additional caveat that though $(P_{\textup{DW}})$ is a LP, it has an exponential number of variables and therefore is unfit for conventional LP solvers. 
We thus adapt the Column Generation Method (CGM), for our particular problem.     

\subsubsection{Solution via Column Generation Method}
The Column Generation Method (CGM) attempts to iteratively solve $(P_{\textup{DW}})$, thereby mitigating the need for directly solving it: starting from $\bGam_0$ - a matrix consisting of a subset of $m_o$ columns of $ \bGam $, one first solves the resulting \emph{restricted master problem} (RMP), i.e. a reduced version of $(P_{\textup{DW}})$. 
Then, at the $l$th iteration, one selects an additional column that is added to $\bGam_0$ (or multiple ones), and solves the resulting RMP. Given a subset $\calX$ of $\calS_\rho$, we define $\bGam(\calX) \in \IZ_+^{U_{\textup{T}} \times |\calX| }  $ as the matrix generated by the $\calX$ columns of $\bGam$, and $\balp(\calX) \in \IR^{|\calX|} $ the corresponding sub-vector of $\balp$.

Let $\calT_o$ be the initial subset of columns for $\bGam$, such that $|\calT_o | = m_o $. 
At iteration $l \geq 1$, given the previously selected columns $\calT_{l-1}$, and the corresponding optimal solutions for the RMP at iteration $l-1$, $\pi_{l-1}^\star $ and $\bmu_{l-1}^\star $,
the vector of \emph{reduced costs} is defined as,
$\bd_l \triangleq \balp(\calZ_{l-1}) - \hat{\bGam}(\calZ_{l-1})^T \bmu_{l-1}^\star - \pi_{l-1}^\star \bone,$ where  
$ \calZ_{l-1} \triangleq \calS_\rho / \calT_{l-1}$ and $ \hat{\bGam}(\calT_{l-1})^T = [ -\bGam(\calT_{l-1})^T , \bI_{U_{\textup{T}}} ]. $ 
Then, the index of the column to be updated is defined as $i_l^\star \triangleq \underset{i \in \calZ_{l-1} }{ \argmin} \ [\bd_l]_i $, 
and the set of \emph{active columns} is updated as follows: $\calT_l = \calT_{l-1} \cup \lrb{i_{l}^\star}$.
Essentially, $i_l^\star$ is the index of the column in $\bGam$, that is added to the RMP.
Then, the updated RMP at iteration $l$ is denoted by $(R_l)$, 
\begin{align} 
(R_l):~~ (\bmu_l^\star , \pi_l^\star) \ \ \begin{cases} 
   \underset{\bmu_l \geq \bzer \ , \ \pi_l }{\argmax} \  \pi_l  ~~~~ \st \ \hat{\bGam}(\calT_l)^T \bmu_l + \pi_l \bone\leq \balp(\calT_l) \     
\end{cases}
\end{align}
where $m_l \triangleq |\calT_l| = m_o + l $, and $ \hat{\bGam}(\calT_l)^T = [ -\bGam(\calT_l)^T , \bI_{U_{\textup{T}}} ]. $
When all reduced costs are non-negative, the optimal solution has been found, i.e., the solution of the current RMP is the same as the original problem. Let $L$ be that iteration number, and $\bw^\star(\calT_L)$, the corresponding RMP solution.
Then, the optimal solution $\bw^\star$ of the original problem, $(P_{\textup{DW}})$ is given by, 
\begin{align} \label{eq:w_opt_CGM}
	\bw_i^\star = 
	\begin{cases}
		\bw_i^\star(\calT_L) \  \textrm{if} \ i \in \calT_L   \\
		 0 , \ \textrm{otherwise}
	\end{cases} 
	, 1 \leq i \leq S  \quad.
\end{align}
Considering~\eqref{eq:w_opt_CGM}, the solution that CGM consists only of the component in $\bw$ that have a contribution to the solution $(P_{\textup{DW}})$, while setting the rest to zero. Interestingly, in most cases, despite the exponential size of $\bw$, it will have only a few non-zero entries.
Note that, in the worst case, CGM ends up adding all columns in $\bGam$, i.e., solving the original problem $(P_{\textup{DW}})$. However, most often, the algorithm will terminate much earlier than that.
Despite its iterative nature, CGM is an exact method, i.e.,  $\bw^\star$ in~\eqref{eq:w_opt_CGM} is the globally optimal solution to $(P_{\textup{DW}})$.
   
\begin{algorithm}
\caption{ DW solution via CGM } \label{alg:CGM} 
\begin{algorithmic}
\State \textbf{Initialization:} $ \calT_0, m_0  $  
\For{$l=1,2, \cdots, S-m_o $}
\State $\calZ_l \leftarrow \calS_\rho / \calT_l $
\State Update $\bGam(\calT_l), \hat{\bGam}(\calT_l), \balp(\calT_l)  $, and solve $(R_l)$ to get $\bmu_l^\star , \pi_l^\star $
\State $\bd_l \leftarrow \balp(\calZ_l) - \hat{\bGam}(\calZ_l)^T \bmu_l^\star - \pi_l^\star \bone   $ 
\State $ i^\star \leftarrow  \underset{i \in \calZ_l }{ \argmin} \ [\bd_l]_i    $
\State \textbf{if} ( $[\bd_l]_{i^\star} \leq 0 $ ) then $\calT_l \leftarrow \calT_l \cup \lrb{i^\star} $
\State \textbf{else} ( $[\bd_l]_{i^\star} > 0 $ ) 
\State \hspace{.2cm} Compute optimal solution in~\eqref{eq:w_opt_CGM}   
\EndFor 
\State \textbf{Output:}   $\bw^\star$
\end{algorithmic}
\end{algorithm}

\subsubsection{Tightness of the DW decomposition}
As a last step, we shed light on the tightness of the proposed decomposition. We derive two simple (yet potentially loose) bounds. 
\begin{lemma}[Bounds on DW decomposition gap] \label{lem:DW_gap}
Let $\bX^\star$ and $\bw^\star$ be the optimal solutions for the primal problem $(P_3)$ and DW problem $(P_{\textup{DW}})$, respectively. Then, 
\begin{align}
0 &\leq f(\bX^\star) - f_{\textup{DW}}(\bw^\star) \leq \eta \sigma_{\max}[\bPsi]  -  \min_{ 1 \leq j \leq S } \alpha_j \leq \eta (\sigma_{\max}[\bPsi]  - \sigma_{\min}[\bPsi] )
\end{align}
where $\eta \triangleq \sum_k \sum_{l \neq k}  \rho_k \rho_l $. 
\end{lemma}
\begin{IEEEproof} Refer to Appendix~\ref{sec:app_dwbound}. \end{IEEEproof}
Interestingly, while the first bound is tighter, the second one is more informative: The DW bound is tighter as the largest and smallest singular values of $\bPsi$ get closer. In the limit case, the DW bound is tight, when \emph{all} the singular values of $\bPsi$ are the same. 

\subsection{Dual Problem } 
We start with deriving the dual problem $(D)$ (a natural lower bound), characterize the resulting duality gap, and compare it to the DW decomposition  bound. In Appendix~\ref{app:dual_prob_anal}, we write the dual problem $(D)$ in a series of equivalent forms, $(D_1)$, $(D_2)$, etc.  Comparing $(D_7)$ in~\eqref{opt:mat_adf_dual_2} to $(P_{\textup{DW}})$ quickly reveals that $(D)$ is a relaxation of $(P_{\textup{DW}})$. Consequently, the bound provided by the DW decomposition is tighter than that of the dual. 
Since the dual problem is the object of several investigations in this section, it is natural to inquire about the wideness of the \emph{duality gap}: the difference between the optimal solution of $(P)$ and that of $(D)$. 
We note that an exact characterization of the duality gap is clearly infeasible (since one needs optimal solutions for $(P)$ and $(D)$). We thus provide a bound on the gap. 
\begin{lemma}[Bound on Duality Gap] \label{lem:duality_gap}
Let $\bX^\star$ and $\blam^\star$ be the optimal solutions for the primal problem $(P_3)$ and the dual $(D)$ in~\eqref{opt:mat_adf_dual}, respectively. Then the duality gap satisfies, 
\begin{align}
0 &\leq f(\bX^\star) - d(\blam^\star) \leq \eta ( \sigma_{\max}[\bPsi] - \sigma_{\min}[\bPsi])  + \bone  \blam^{\star} - \sum_k \rho_k \ \min_i \ [\blam^\star]_i
\end{align}
where $\eta \triangleq \sum_k \sum_{l \neq k}  \rho_k \rho_l $. 
\end{lemma}
\begin{IEEEproof}
Refer to Appendix~\ref{app:duality_gap_proof}
\end{IEEEproof}
Unlike the bound on the DW gap (Lemma~\ref{lem:DW_gap}), one cannot draw informative conclusions about potential limit cases, where the duality gap might collapse. 

\paragraph*{Section Summary} 
Motivated by the lack of optimality claims on the BCD solution, we derived lower bounds such as the DW decomposition, and the dual problem. 
We showed that the dual problem is a relaxation of the DW problem. Consequently, the DW problem offers as good a bound as possible (or better) with respect to the dual problem. This in turn implies that methods based on the DW decomposition (e.g., CGM) yield tighter approximations than methods based on the dual problem (e.g., dual subgradient ascent, Lagrange relaxation). 

\section{Numerical Evaluations} \label{sec:numres}
We present in the following a performance evaluation of our proposed algorithm. The evaluation is based on simulations, and in addition to comparing our proposed approach to some benchmark schemes, we also evaluate the presented bounds for the leakage.

Recalling the discussion in Sec.\ref{sec:problem_statement}, we note that achieving good sum-rate performance is intimately linked with interference management. We will provide empirical evidence underlining the severe limitations of exhaustively searching for precoder combinations, that  maximize the sum-rate, in Remark~\ref{rmk:exh_ser} (below). 
That being said, W-MMSE~\cite{shi_wmmse_2011} is one of the few algorithms that guarantee locally optimal solutions (after large enough number of iterations). By including Weighted-MMSE in the numerical results, we will provide insight into the relation between the (locally) optimal sum-rate that can be achieved in a given network, and the sum-rate achieved by our leakage-based scheme.

\begin{remark}[Exhaustive Search for Optimal Precoder] \label{rmk:exh_ser} \rm 
We attempted to find globally optimal sum-rate solutions (by quantizing the precoders at each antenna main, and exhaustively searching for the optimum), for a small system with $ A = 2, M=4, N=2, U_{\textup{T}} = 16 $. 
We set a reasonable number of precoder choices, to yield sum-rate values that are close to the true optimum. Then, getting the optimal sum-rate, for each Monte Carlo point, was taking $\sim 20$ hours. Thus, simulating a $100$ Monte Carlo points would be clearly infeasible. 
Similar limitations were reported in~\cite{Mochaourab_Benders_15}, where the difficulty of exhaustively searching of a toy example (one BS with $2$ antennas and $2$ single antenna users), was underlined. 
These findings are direct consequences of the NP-hard nature of the sum-rate maximization problem, which renders exhaustive search too complex even for small toy examples.  
\end{remark}

\subsection{Methodology}
Recall that $A$ is the total number of antenna domains, $N$ and $U$ the number of radio-heads and users per antenna domain, respectively, and $M$ the number of antennas at each radio-head. 
Central processors, radio-heads and users are positioned within the area of interest, of size $A\Delta^2$, $\Delta=100$m. 
While positions of central processors/radio-heads are fixed throughout the simulation, users are dropped uniformly for each simulation realization (users are static and mobility is only considered in Sec.~\ref{sec:sr_wmob}). 
Averaging is done over $100$ different independent such simulation realizations.
To emulate LoS propagation in C-RAN, channels between radio-heads and users are assumed to be spatially correlated Rician (Kronecker model), with pathloss and shadow fading. The parametrization is discussed at length in~\cite{Rahman_RRHclust_15}[Sect. VII-A]. 
Moreover, the following schemes are investigated.  
\begin{itemize}
\item[o] \emph{Proposed UA method:} Its operation is detailed in Sec~\ref{sec:syslev_oper}. 
Note that several choices exits for the precoder design, such as zero forcing precoder, leakage minimizing precoder, etc..   
Throughout the numerical results, the initial precoders are selected as zero-forcing (with equal power allocation among all users), to null all intra-AD interference (outlined in Appendix~\ref{prop:opt_prec}). 
Thus, the precoder \emph{is not} iteratively updated within the proposed algorithm, but rather fixed (it is chosen to be zero forcing, though many other choices exist). 
\end{itemize}
We benchmark our proposed UA method against the following schemes:
\begin{itemize}
\item[o] \emph{Distance-based assignment:} Users are associated to radio-heads (and consequently antenna domains) based on strongest channels ($\rho_i$ users are associated to antenna domain $i$). Finally, each antenna domain performs zero forcing to its users. 
\item[o] \emph{W-MMSE (without UA):} The set of assigned users is (randomly) selected in advance, and fixed (no user assignment is performed). Then $10$ iterations of W-MMSE are carried out, to optimize the preocoders. 
\item[o] \emph{ILM (without UA):} Interference Leakage Minimization (ILM), by finding the well-known leakage minimizing precoders~\cite{gomadam_distributed_2011}. It is a special case of our proposed method, where assigned users are selected in advance and fixed and only precoding is done. The set of assigned users is the same for W-MMSE and ILM.  
\end{itemize}

\subsection{Sum-rate Results}
\begin{figure}
  \begin{center}
  \begin{minipage}{7cm}
    \centering
    \includegraphics[trim={0 0 0 1.7cm}, clip, height=6cm, width=8cm]{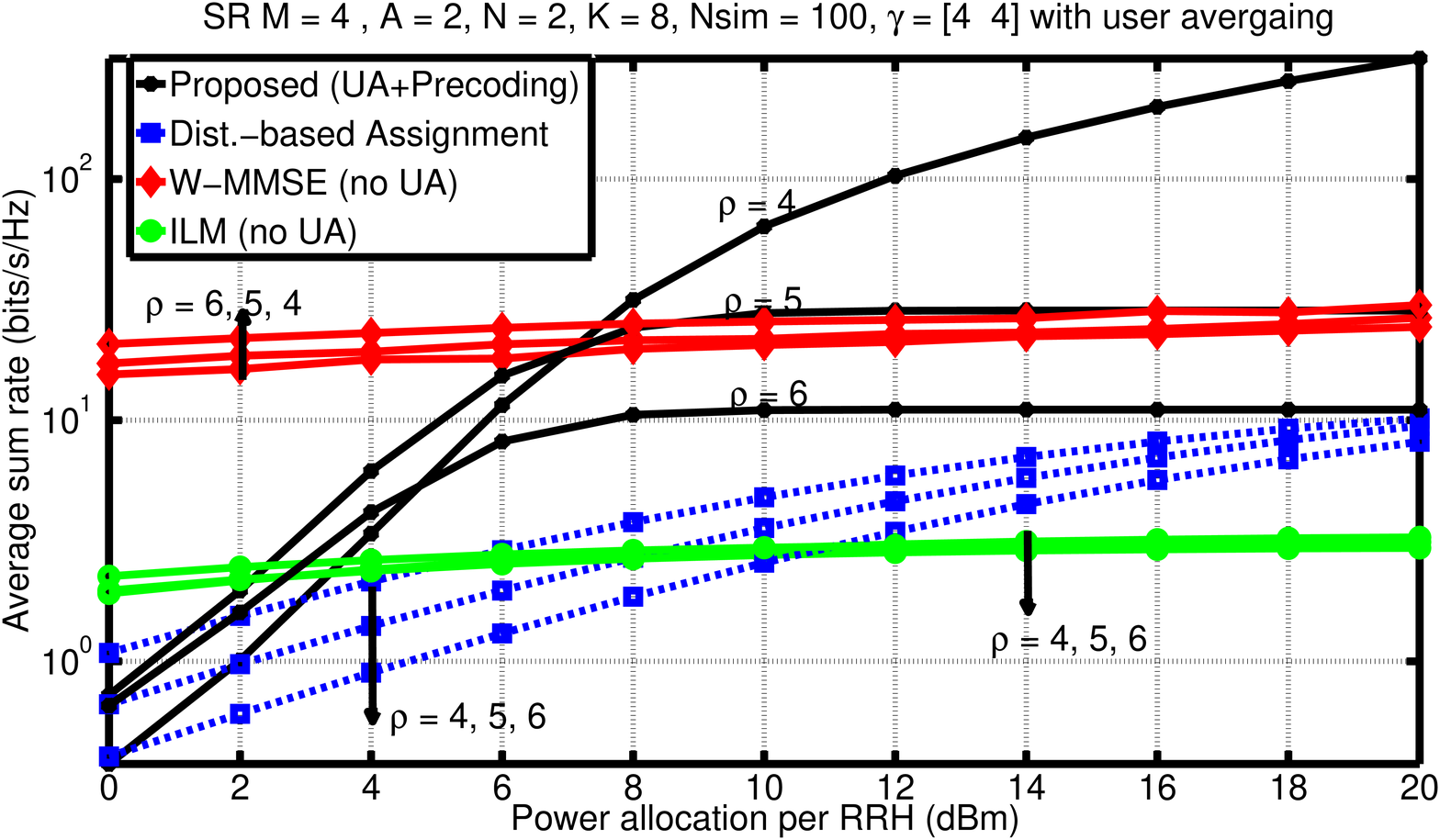}
    \caption{Average sum-rate performance \emph{in log scale}, for $A = 2, M=4, N=2, U=8, U_{\textup{T}} = 16 $ }
  \label{fig:sr_small}
  \end{minipage}
  \hspace{1cm}
  \begin{minipage}{7cm}
    \centering
    \includegraphics[trim={0 0 0 1.7cm}, clip,height=6cm, width=8cm]{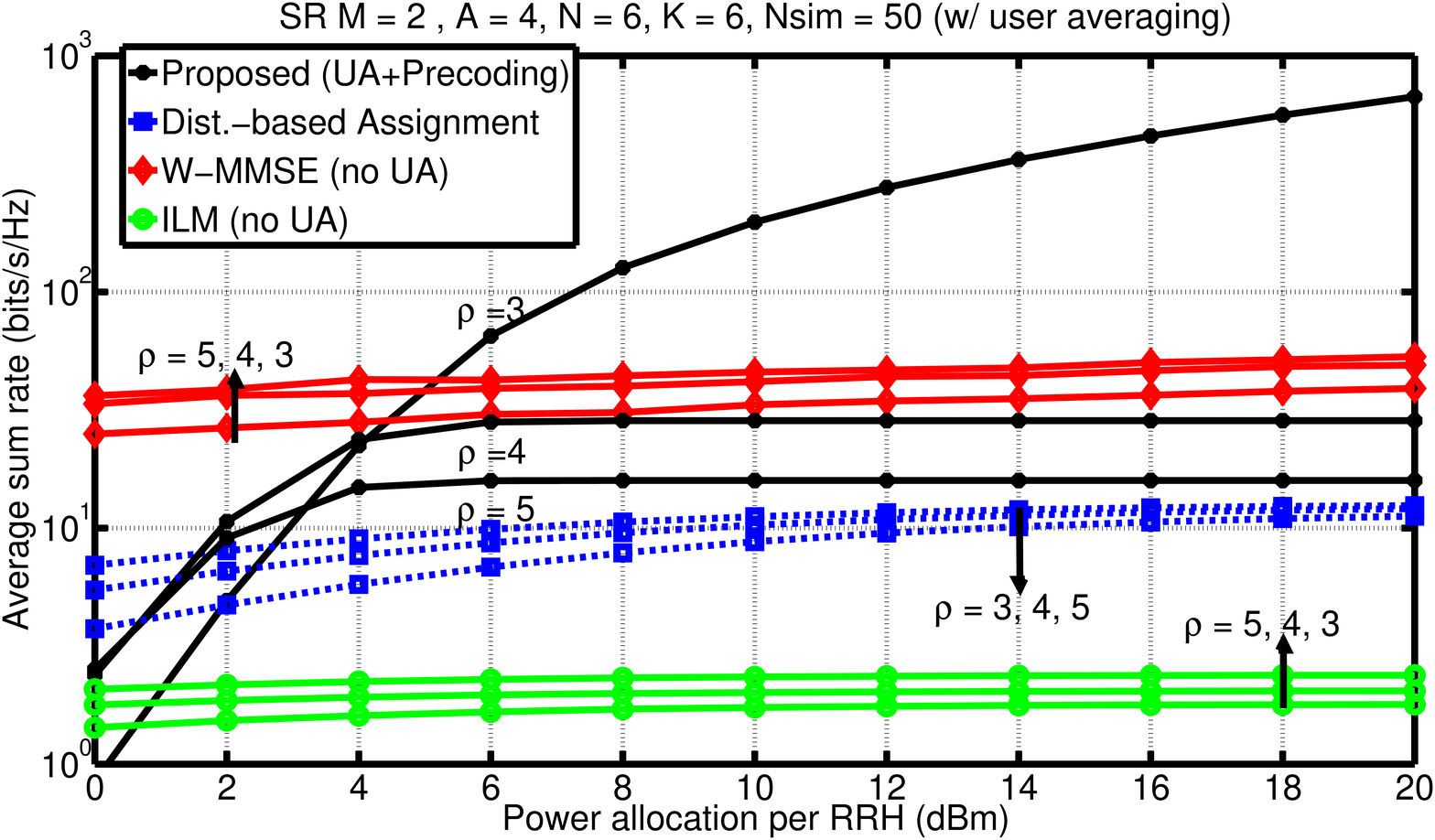}
      \caption{Average sum-rate performance, \emph{in log scale} for $A=4, M=2, N=6, U=6, U_{\textup{T}} = 24 $ }
  \label{fig:sr_large}
  \end{minipage}
  \end{center}
\end{figure}

We first aim to investigate the sum-rate performance of a relatively small deployment with $A = 2, M=4, N=2$ radio-heads per antenna domain, and $U=8$ users per antenna domain (for a total of $U_{\textup{T}} = 16$ users), while varying the loading factors $\rho$. Note that sum-rate values are plotted in \emph{log scale}, for clarity. 
Fig.~\ref{fig:sr_small} shows the resulting sum-rate, and one can clearly see an increase in the performance of all schemes, as $\rho$ is decreased. 
This result is expected since interference decreases as less users are served. 
More importantly, we see a very significant performance gap between our proposed method, and both the distance-based and ILM benchmarks (recall that ILM is nothing but our proposed scheme, without any UA). Moreover the aforementioned gap is increasing with decreasing $\rho$.  
We also underline that the performance of our scheme is close to that of W-MMSE, for $\rho=5,6$ in the medium/high SNR regime, and significantly outperforms it when $\rho=4$. The latter is due to the fact that the proposed scheme can completely suppress all interference (refer to the discussion below). 

In addition, it is well established that schemes optimizing the interference leakage, are suboptimal in the very low-SNR region. This is the reason behind the poor low SNR performance of the proposed, ILM, and distance-based schemes (refer to~\cite{Ghauch_IWU_15} for a more detailed discussion). We notice as well a rather quick degradation in the sum-rate of our proposed scheme, as the loading factor increases. This sensitivity can be attributed to the specific choice of using zero-forcing as precoding, in the numerical results.
This does suggest that a careful choice of the loading factors is needed, to avoid heavily-loaded cases, i.e., $\rho = 6$ in Fig.~\ref{fig:sr_small}. Despite this shortcoming,  the proposed scheme still offers a significant improvement in sum-rate performance, over the distance-based and ILM benchmarks (as sum-rate values displayed in log-scale).

Similar trends are observed by moving on to a larger setup where $A = 4, M=2, N=6$ radio-heads per antenna domain, and $U=6$ users per antenna domain (for a total of $U_{\textup{T}} = 24$ users), as evidenced in Fig.~\ref{fig:sr_large}. 
However, we clearly see that in this case, the performance gap is indeed more pronounced than that of the previous case (Fig.~\ref{fig:sr_small}): While the performance of the benchmark increases with smaller $\rho$, this increase is significantly more pronounced for our algorithm. In particular, for the case where $\rho=3$, the gap is significant.

In the heavily loaded case (i.e., $\rho = 6$ in Fig.~\ref{fig:sr_small} and $\rho = 5$ in Fig.~\ref{fig:sr_large}), the performance of the distance-based assignment approaches that of our proposed scheme. We note that this only happens after $20$ dBm, which is already on the upper limit of RRH transmit power. Moreover, for realistic operating points, e.g., $15$ dBm, our scheme has almost twice the sum-rate performance of that same benchmark in Fig.~\ref{fig:sr_large} (in fact, that gap is still $2$ bps/Hz at $20$dBm). We recall that this follows from the sum-rate values being displayed in log-scale. 

\begin{figure}
  \begin{center}
  \begin{minipage}{7cm}
    \centering
   \includegraphics[trim={0 0 0 1.7cm}, clip,  height=6cm, width=8cm]{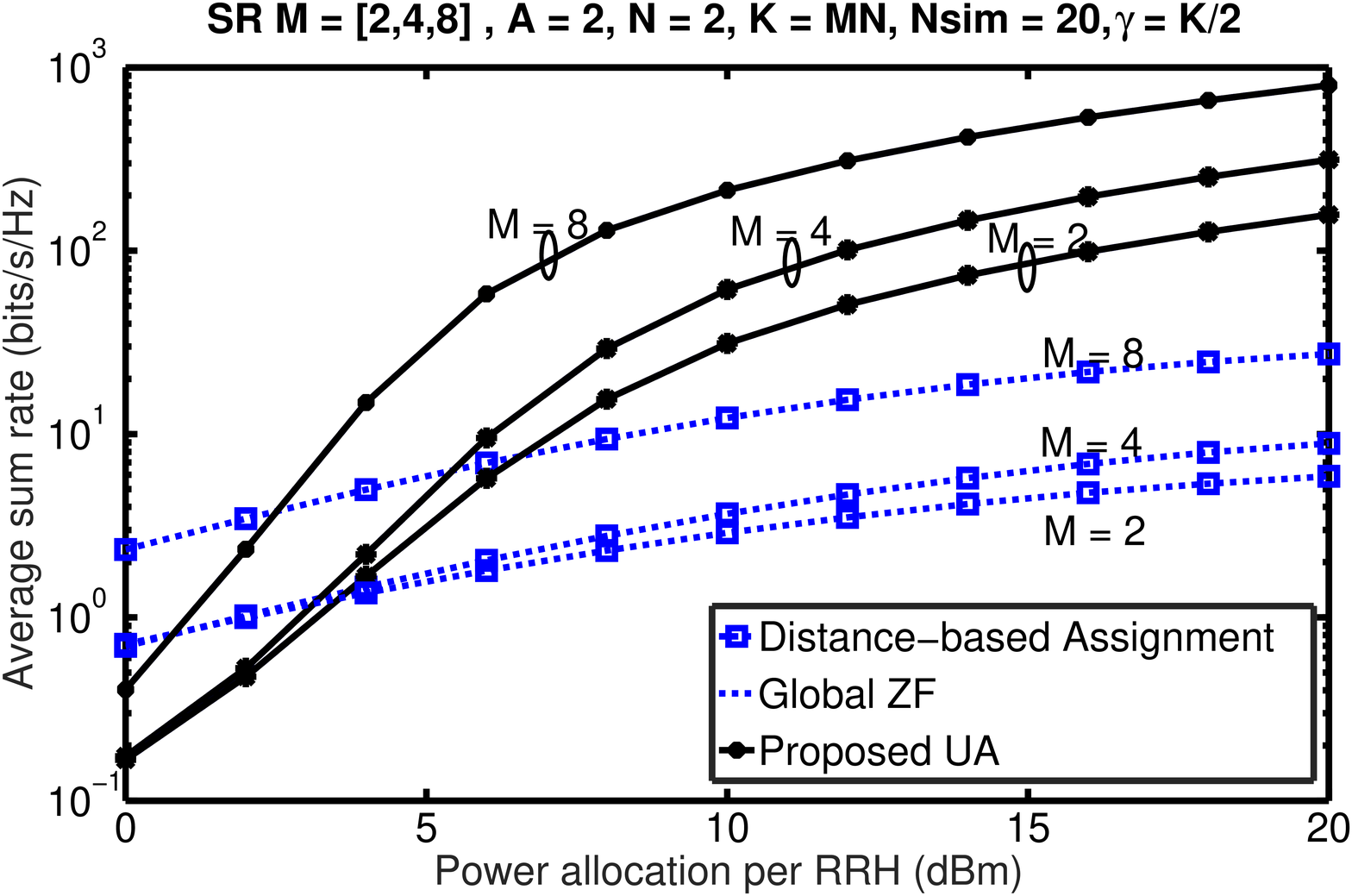}
   \caption{Average sum-rate performance for $A=2, U=MN, \rho= U/2 $ (curves for Proposed UA and Global ZF coincide) }
     \label{fig:sr_UB}

  \end{minipage}
  \hspace{1cm}
  \begin{minipage}{7cm}
    \centering
    \includegraphics[ height=5cm, width=7cm]{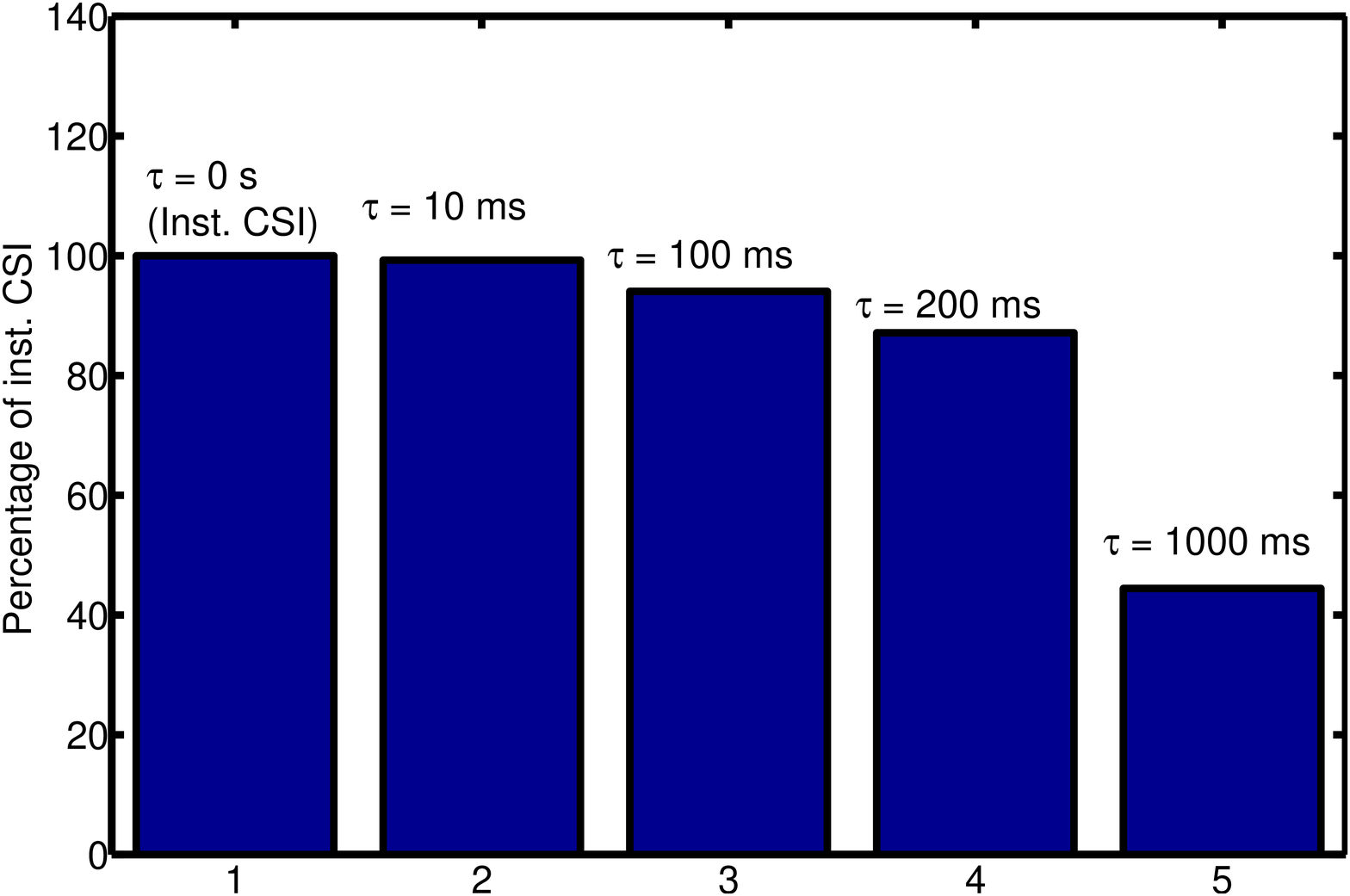}
    \caption{ Sum-rate degradation for several values of $\tau$ ( SNR$=12$dB , $A = 2, M=4, N=2, U=8, U_{\textup{T}} = 16, \rho = 4 $  }
  \label{fig:sr_small_wmob}
  \end{minipage}
  \end{center}
\end{figure}

We next investigate some specific choices of loading factors, i.e., deployments with $A=2$, $N=2$, $U=MN$, and where the loading factor is appropriately chosen as $\rho = U/2 $. 
In this case, we also consider the performance of global zero-forcing, as an upper bound on the system performance.
Fig.~\ref{fig:sr_UB} shows the sum-rate for such a system, for various values of $M$. 
Most importantly, in this regime, our proposed algorithm coincides exactly with that of the global ZF upper bound. 
This is due to the fact that in this case our scheme is able to totally suppress all interference in the network. 
This finding can be generalized in the following way: Regarding the choice of loading factors, when $\sum_i \rho_i \leq MN $ then the leakage can be completely nulled. 
One can see this by considering a special case of the precoder design outlined in Appendix~\ref{prop:opt_prec}, by replacing $U$ with $\rho_i$. It this case, it is straightforward to show that the interference leakage is zero, when $\sum_i \rho_i \leq MN $.
This explains the above observed result where all interference is completely nulled, turning the system into a (virtually) noise-limited one.    

\subsection{Effect of Mobility and Outdated CSI} \label{sec:sr_wmob}
We next address the issue of user mobility and lack of instantaneous CSI, when central processors employ \emph{outdated CSI} (for both the UA algorithm and precoding), and low user mobility: what is the performance loss associated with using CSI, from previous scheduling time-slots?
In this case we have used the same simulation scenario as that of Fig~\ref{fig:sr_small}, under a simple user mobility model (each user picks up a random direction of motion between $0 - ~2\pi$, and random speed between $0 - 40$ Km/h). The average SNR across all users is fixed to $12$dB, and the degradation (in percent) is measured w.r.t. the instantaneous CSI case. 
Fig~\ref{fig:sr_small_wmob} suggests virtually no performance degradation when outdated CSI is between $1$ and $10$ ms, and negligible degradation when using CSI from $100$ms ago. This results in $2-100$ times reduction in CSI acquisition overhead, with little-to-no performance loss.

\subsection{Bounds on Interference Leakage }
We next compare the performance of the proposed BCD algorithm (Algorithm~\ref{alg:1}) against the globally optimal solution (found via exhaustive search), as well as the DW lower bound. 
We first look at the tightness of the DW decomposition, with respect to the globally optimal solution of $(P)$. We consider a small scenario ($A=2$), assuming no fading, and looking at the (average) total interference leakage $f$, as metric. 
As seen in Table~\ref{table:bcd_vs_DW}, the error from approximating the globally optimal solution of $(P)$, by the DW lower bound (solved using CGM in Table~\ref{alg:CGM}) is quite tolerable, for $\rho = 3, 4$ (the case where $\rho=1$ is not practically relevant).
These results suggest that the DW bound better approximates the solution to the UA problem, as the dimensions of the system grow. 
However, verifying this hypothesis is challenging due to the exhaustive search step needed for solving $(P)$.   
We also compare in Table~\ref{table:bcd_vs_DW} the performance of the proposed BCD algorithm (Algorithm~\ref{alg:1}) against that of the globally optimal solution. 
We observe a similar trend here, where the proposed BCD algorithm has a similar performance as the globally optimal solution, for relevant cases. 

\begin{table} 
\begin{center}
\begin{tabular}{ |c|c|c|c|}
 \hline
       &  $ N=1, U=2, \rho=1$ & $N=2, U=4, \rho=3$ & $ N=3,U=6, \rho=4 $  \\
 \hline        
Prop.  &  $ 0.5329$       &  $7.5445$    & $12.1334$ \\
Primal Opt  &  $ 0.3443$       &  $ 6.7538$    & $10.8226$   \\ 
DW decomp      &  $ 0.2392$     &  $5.9249$    & $ 9.3255$     \\ 
\hline
\textbf{Error (DW)} ($\% $) & $\pmb{ 30.53 }$     &  $ \pmb{13.99} $       & $ \pmb{16.05}$     \\ 
\textbf{Error (Prop)} ($\% $) & $\pmb{ 54.78 }$     &  $ \pmb{11.71} $       & $ \pmb{12.11}$     \\ 
 \hline
\end{tabular} 
\end{center}
\caption{\label{table:bcd_vs_DW} Average total interference leakage: proposed algorithm vs DW lower bound vs globally optimal, for $A=2, M=2, U=MN $  }
\vspace{-1em}
\end{table}

\subsection{Communication Overhead and Complexity } \label{sec:comm_overhead}
At each central processor, the \emph{computational complexity} of the proposed method (described in Sec.~\ref{sec:syslev_oper})  is dominated by the matrix inversion ($MN\times MN$) (refer to Appendix~\ref{prop:opt_prec}), and solving~\eqref{opt:xk_opt}. 
The resulting complexity at each central processor is approximately $\calC_{\textup{prop}} =  (MN)^{2.5} + U_{\textup{T}}^3$. 
Though intended to be distributed, W-MMSE can be deployed in a C-RAN setup. 
Its complexity at each central processor can be approximated as
$\calC_{\textup{wmmse}} =  L_{\textup{WMMSE}}(MN)^{2.5}  $, where $(MN)^{2.5}$ is due to matrix inversion\cite{shi_wmmse_2011}, and $L_{\textup{WMMSE}}$ is the number of iterations. Thus, it is clear that $\calC_{\textup{wmmse}}/\calC_{\textup{prop}} \approx L_{\textup{WMMSE}} $, as $L_{\textup{WMMSE}} $ increases, implying a huge gap in complexity (e.g., an order-of-magnitude for $10$ W-MMSE iterations). Moreover, it is known that the algorithm is slow in convergence~\cite{Schmidt_comparison_13}.    
This becomes even more critical since W-MMSE requires more iterations until convergence, as the system dimensions grow. Thus, its scalability is severely limited which is a critical issue in densely deployed networks like C-RAN.

Moreover, we (roughly) estimate the cost associated with deploying the proposed method (described in Sec.~\ref{sec:syslev_oper}), in terms of \emph{total communication overhead}. We use the coarse measure of counting the total number of required \emph{training symbols}. 
The overhead of the proposed approach chiefly consists of UA overhead in Algorithm~\ref{alg:1} (totaling $ A L_{\textup{UA}} (U_{\textup{T}}/8)~\textrm{symbols} $, $L_{\textup{UA}}$ is the number of iterations for Algorithm~\ref{alg:1})\footnote{We assume that each assignment vector $\bx_k$ is encoded into $8$-bits.}, the CSI acquisition overhead (amounting to $  U_{\textup{T}} A M N $  symbols), the data sharing overhead (equaling $AU = U_{\textup{T}}  ~\textrm{symbols}$), and the radio-head synchronization overhead (amounting to $ AU = U_{\textup{T}} ~\textrm{symbols}$).    
Using the above reasoning, we note that distributed coordination algorithms such as W-MMSE can be run in a C-RAN context. The resulting overhead for W-MMSE is similar to that of the proposed scheme.

\subsection{Discussions} \label{sec:disc}
A clear observation that follows from the above results (Fig.~\ref{fig:sr_UB}), is that huge performance gains can be achieved when the loading factors are appropriately chosen. 
Though the performance of our proposed scheme can be close to that of global ZF, it circumvents the corresponding need for synchronizing all radio-heads in the system. Not surprisingly, we observe that the performance depends on $MN$, the total number of transmit antennas in each antenna domain, rather than on $M$ and $N$, individually. 
We reiterate the fact that the performance of our proposed approach depends on the initial state of the network (a direct consequence of our definition for the UA problem).
Our results also suggest that both the proposed BCD-based algorithm (Algorithm~\ref{alg:1}), and the DW lower bound approximate well the globally optimal solution to the UA problem, for practical cases. 
With that in mind, solving the DW problem via the low-complexity CGM provides an efficient means of estimating (optimistically) the residual interference in the network, after applying the proposed method.


\section{Conclusions}
We formulated the UA problem in a multi-AD C-RAN as an integer optimization problem (using the interference leakage as metric), and showed that it can be efficiently solved by a BCD scheme. 
Motivated by lack of optimality guarantees on the BCD solution, we argued the need for `good' lower bounds on the problem (i.e., the total interference leakage).  
We investigated several classical lower bounds and showed that the DW problem offers tighter bounds than the dual problem, and adapted the Column Generation Method to (globally) solve the DW problem. 
Our numerical results showed that the proposed UA algorithm is within a $20\%$ sum-rate gap compared to W-MMSE, however with ten times lower complexity: It offers a scalable alternative to W-MMSE-type approaches which are ill-suited to operate in densely deployed C-RANs. We also observed that it outperforms W-MMSE, under some specific loading conditions.  
Finally, the proposed UA algorithm and the DW lower bound seem to approximate the optimal solution to the UA problem, with acceptable error, in practical setups. 

\section{Acknowledgment}
The authors are grateful to the editor and anonymous reviewers, for all their efforts in improving the manuscript. 

\appendix
\section{}
\subsection{Proof of Proposition~\ref{prop:prob_form}  } \label{sec:app_prob_form} 
The fact that $(P)$  can be rewritten in vector form, i.e., $(P_2)$, is straightforward and can be skipped. 
As for rewriting $(P)$ in matrix form, i.e., $(P_3)$,   
we first recall that for any $\bQ \in \IR^{m \times m} $, $ \pmb{1}^T \bQ \pmb{1}  = \sum_{i=1}^m \sum_{j=1}^m Q_{i,j}  $, and rewrite the cost function in $(P)$ as, 
\begin{align}
f &= \tr[ \pmb{1}^T ( \bX^T \bPsi \bX ) \pmb{1} ] - \tr( \bX^T \bPsi \bX ) = \tr(   \bX^T \bPsi \bX  \pmb{1} \pmb{1}^T   ) - \tr( \bX^T \bPsi \bX ) = \tr \left( \bX^T \bPsi \bX \pmb{\Omega} \right)
\end{align}
where we used the fact that $\tr(\bA \bB) = \tr(\bB \bA) $, and let $\pmb{\Omega} \triangleq  \pmb{1} \pmb{1}^T - \bI_A  $. Moreover, the loading constraint can be rewritten as,
$\sum_{i_m} x_{k,i_m} = \rho_k , \ \forall k \Leftrightarrow  \pmb{1}^T \bX = [\rho_1, ..., \rho_A ] \Leftrightarrow \bX^T \pmb{1} = \pmb{\rho}.  $
The assignment constraint can be reformulated as,
 $\sum_{k=1}^A x_{k,i_m} \leq 1, \forall i_m \in \calU_{\textup{T}} \Leftrightarrow \bX \pmb{1} \leq \pmb{1}.  $
\vspace{-1.5em}

\subsection{Precoding Design for Numerical Evaluations } \label{prop:opt_prec}
We provide some guidelines for the precoder design needed for the numerical evaluations in Sec.~\ref{sec:numres}.  For convenience, we let
$\bH_{i,j} \triangleq
\begin{bmatrix}
\bh_{i,j_1}^T , ~\cdots, ~\bh_{i,j_U}^T 
\end{bmatrix}^T$, $\bH_{i,j} \in \IC^{U \times MN}$, and 
$\bV_i \triangleq [ \bv_{i_1} , \cdots , \bv_{i_U} ] $, $\bV_{i} \in \IC^{MN \times U}$,
denote the channel between the antennas of antenna domain $i$ and the users of antenna domain $j$, and the matrix of precoding vectors for antenna domain $i$, respectively. 
This precoder is defined as, 
\begin{align} \label{opt:prec_opt}
\bV_i^\star = \begin{cases} 
   \underset{ \bV_i \in \IC^{MN \times U} }{\argmin} ~~ \tr( \bV_i^\dagger \bR_i \bV_i ) \ \ \st \ \bH_{i,i} \bV_i = \beta_i \bI_U \\
\end{cases}
\end{align}
where $ \bR_i \triangleq \sum_{ \substack{ j \in \calA \\ j \neq i}}  \bH_{i,j}^\dagger \bH_{i,j}$, and $\beta_i$ is a design parameter to fulfill a maximum transmit power. 
Note that in this specific design choice, intra-AD is nulled (following the constraint in~\eqref{opt:prec_opt})
The precoder is the optimal solution to the above, and is given by, 
\begin{align}
&\bV_i^\star = { \sqrt{U} \ \bR_i^{-1} \bH_{i,i}^\dagger \left(  \bH_{i,i} \bR_i^{-1} \bH_{i,i}^\dagger \right)^{-1} }{ / \Vert \bR_i^{-1} \bH_{i,i}^\dagger \left(  \bH_{i,i} \bR_i^{-1} \bH_{i,i}^\dagger \right)^{-1}  \Vert_F } .
\end{align}
Moreover, for $U \leq MN $, the problem is feasible almost surely.

\vspace{-1.5em}

\subsection{Proof of Lemma~\ref{lem:mono} } \label{sec:app_mono}
Note that the following is a direct consequence of~\eqref{opt:xk_opt} 
\begin{align*}
f( \lrb{\bx_k^{(n)}} ) &\geq f(\bx_1^{(n+1)}, \bz_1^{(n)}  )  \geq f(\bx_2^{(n+1)}, \bz_2^{(n)}  ) ... \geq f(\bx_A^{(n+1)}, \bz_A^{(n)}  ) \triangleq f( \lrb{\bx_k^{(n+1)}} )   
\end{align*}
where the last equality follows from the fact that $f(\bx_A^{(n+1)}, \bz_A^{(n)}  )$ corresponds to the case where all variables $(\bx_1, ...., \bx_A)$, are updated. It follows that the sequence $ \lrb{ f(\bx_1^{(n)}, ..., \bx_A^{(n)} ) }_n $ converges to a limit point $f_o$. 


%

\subsection{Proof of Lemma~\ref{lem:DW_gap} } \label{sec:app_dwbound}
Let $\eta = \sum_k \sum_{l \neq k} \rho_k \rho_l $.
The left inequality follows immediately from the fact that the DW decomposition is always a lower bound on the problem - by construction (Sec~\ref{sec:DWD}).  
 Moreover, the right one is obtained from upper bounding $f(\bX^\star)$ and lower bounding $f_{\textup{DW}}(\bw^\star)$, 
\begin{align*}
f(\bX^\star) &= \sum_k \sum_{l \neq k} \bx_k^{\star^T} \bPsi \bx_l^\star \leq \sum_k \sum_{l \neq k} \sigma_{\max}[\bPsi] \Vert \bx_k^\star \Vert_2 \Vert \bx_l^\star \Vert_2  \overset{(d.1)}{=} \sigma_{\max}[\bPsi] \sum_k \sum_{l \neq k} \rho_k \rho_l = \sigma_{\max}[\bPsi] \eta
\end{align*}
where $(d.1)$ follows from the fact that $\bx_k^\star$ must be feasible: thus, $\Vert \bx_k^\star \Vert_2$ is the sum of all non-zero elements, and equal to $\rho_k$. Moreover, a simple/naive lower bound can be obtained on $P_{\textup{DW}}$ by relaxing the first constraint, 
\begin{align*}
f_{\textup{DW}}(\bw^\star) \geq 
\underset{ \substack{ \bone^T \bw = 1, \\ \bw \geq \bzer   }  }{\min} \  \balp^T \bw \overset{(d.1)}{=} \min_{1 \leq j \leq S } \ \alpha_j = \min_j \ \tr(\bQ_j^T \bPsi \bQ_j \pmb{\Omega} )  \overset{(d.2)}{\geq} \eta \sigma_{\min}[\bPsi]
\end{align*}
where $(d.1)$ follows from the fact that problem is a special LP, whose solution is obtained in Definition~\ref{sec:app_specLP}. 
Moreover, $(d.2)$ follows similar reasoning used for lower bounding $d(\blam^\star)$ in Appendix~\ref{sec:app_dwbound}. 
The first and second bound follows from combining $(d.1)$ and $(d.2)$ respectively. 
\vspace{-1em}
\subsection{Dual Problem Analysis} \label{app:dual_prob_anal}
The dual problem, $(D)$, is defined as,
\begin{align} \label{opt:mat_adf_dual}
(D)\underset{\blam \geq \bzer }{\max} d(\blam) = \lrb{ \underset{\bX \in \calS_\rho }{\min} \tr ( \bX^T \bPsi \bX \pmb{\Omega}  )  + \blam^T(\bX \pmb{1} - \pmb{1} ) }  
\end{align}
and written equivalently as (some steps are omitted due to limited space),
\begin{align*}
&(D_5) \ \underset{ \blam , \zeta }{\max} \ \zeta \ \ \st \ \balp  + \bGam^T \blam \geq \zeta \bone , \ \blam \geq \bzer_{U_{\textup{T}}}
\end{align*}
Letting $\bmu = [\blam , \ \zeta  ]^T $, $\bc = [ \bzer , \ 1 ]^T $, and $\bar{\bGam}^T = [ -\bGam^T , \  \bone] $, $(D_5)$ is equivalent to, 
 \begin{align} \label{opt:mat_adf_dual_1}
 (D_6) \ \underset{ \bmu  }{\max} \ \ \bc^T \bmu \ \ \st \  \bar{\bGam}^T \bmu \leq \balp  ,  \ \bmu \geq \bzer
\end{align}
where  $\bc = [ \bzer, \ 1 ]^T $, and $\bar{\bGam}^T = [ -\bGam^T , \  \bone ] $.  
The above problem is a LP, and since strong duality holds, we work with its (equivalent) dual form. 
Pluging in the values of $\bar{\bGam}$ and $\bc$, $(D_6)$ becomes,
\begin{align}  \label{opt:mat_adf_dual_2}
(D_7) \ \underset{\bw}{\min} \ \balp^T \bw   \ \  \st \ \bGam \bw \leq \bzer, \ \bone^T \bw \geq 1 , \  \bw \geq \bzer    
\end{align}
\vspace{-2em}
\subsection{Proof of Lemma~\ref{lem:duality_gap}  } \label{app:duality_gap_proof}
Let $\eta = \sum_k \sum_{l \neq k} \rho_k \rho_l $.
The left inequality follows immediately from weak duality. Moreover, the right one is obtained from upper bounding $f(\bX^\star)$ and lower bounding $d(\blam^\star)$. The upper bound on $f(\bX^\star)$  follows the same reasoning as that of Appendix~\ref{sec:app_dwbound}, and yields $f(\bX^\star) \leq \sigma_{\max}[\bPsi] \eta$. 
Using the dual problem in~\eqref{opt:mat_adf_dual},  we formulate the optimal dual solution (and its lower bound), 
\begin{align*}
d(\blam^\star)  &= \underset{ \substack{ \bx_k \in \IB^{U_{\textup{T}}} , \forall k  \\ \bx_k^T \bone = \rho_k , \forall k } }{\min} \sum_k \bx_k^{T}  ( \sum_{l \neq k} \bPsi \bx_l + \blam^\star ) - \bone^T \blam^\star \\
&\geq \underset{ \substack{ \bx_k \in \IB^{U_{\textup{T}}} , \forall k  \\ \bx_k^T \bone = \rho_k , \forall k } }{\min} \sum_k \left(  \sum_{l \neq k} (\sigma_{\min}[ \bPsi ] \Vert \bx_k \Vert_2   \Vert \bx_l  \Vert_2 ) + \bx_k^T \blam^\star \right) - \bone^T \blam^\star \\
&\overset{(f.2)}{=} \sigma_{\min}[\bPsi] \eta  - \bone^T  \blam^{\star} +\sum_k \underset{ \substack{ \bx_k \in \IB^{U_{\textup{T}}}   \\ \bx_k^T \bone = \rho_k  } }{\min} \bx_k^T \blam^\star  \overset{(f.3)}{=} \sigma_{\min}[\bPsi] \eta  - \bone^T  \blam^{\star}
+\sum_k \underset{ \substack{ \bx_k \geq \bzer  \\ \bx_k^T \bone = \rho_k  } }{\min} \bx_k^T \blam^\star \\
&\overset{(f.4)}{=} \sigma_{\min}[\bPsi] \eta  - \bone^T  \blam^{\star}
+\sum_k \rho_k \big( \underset{ \substack{ \bz_k \geq \bzer  \\ \bz_k^T \bone = 1  } }{\min}  \bz_k^T \blam^\star \big) \overset{(f.5)}{=} \sigma_{\min}[\bPsi] \eta  - \bone^T  \blam^{\star} + \sum_k \rho_k \ \min_i \ [\blam^\star]_i
\end{align*}
Note that  $(f.2)$ follows from the fact that $\Vert \bx_k \Vert_2 = \rho _k$ for any feasible $\bx_k$.
$(f.3)$ is due to the fact that the problem is a MILP. Furthermore, we show that it satisfied the integrality property (as per Definition~\ref{def:integrality}): then, relaxing the binary constraint into a continuous one, yields the optimal solution. Finally, $(f.4)$ is obtained by letting $\bz_k = \bx_k /\rho_k $, and $(f.5)$ from the fact that the problem is a Special LP whose solution is detailed in Definition~\ref{sec:app_specLP}. 
The final result follows by combining the above result with $(d.1)$. 
\vspace{-1em}


\subsection{The Two Antenna Domain Case} \label{sec:special_cases}
We focus in this section on the case of two antenna domains.
Firstly, the cost function is given by $f(\bx_1, \bx_2 ) =   \bx_1^T (\bPsi + \bPsi^T )\bx_2 \triangleq \bx_1^T \bar{\bPsi} \bx_2 $, and the assignment constraint is always satisfied (it can be dropped). Assuming full-load conditions with equal loading (i.e.,  $\rho_1 = \rho_2 = U_{\textup{T}}/2$), the relation $\bx_2 = \bone -  \bx_1  $, can be used to express the UA problem in terms of $\bx_1$ only: we thus drop all subscripts, and the loading constraint becomes, $\bone^T \bx = \rho $, and $\calS_{\rho } = \lrb{ \bx \in \IB^{U_{\textup{T}}}  \ | \ \bone^T \bx = \rho  } $. $(P_2)$ takes the following simple form, 
\begin{align} \label{opt:UA_2AN}
 (P_4): \ f(\bx^\star) = \underset{ \bx \in \calS_{\rho} }{\min} \ f(x) = (\bx^T \bar{\bPsi} \bone- \bx^T \bar{\bPsi} \bx )  
\end{align}

We use a `DW-like' transformation to reformulate $(P_4)$, into an equivalent form, using Lemma~\ref{lem:opt_DWD} (that is shown below). Applying it to $(P_4)$ yields,
\begin{align*}
(P_5) \  \bw^\star =
\begin{cases}
 \argmin \ \bw^T \balp \ \ \ \st  \ \bw^T \bone = 1, \bw \geq \bzer
\end{cases}
\end{align*}
where  $\balp = [ \alpha_1 , ...., \alpha_S ]^T $, $\alpha_j = \bu_j^T \bar{\bPsi} \bone - \bu_j^T \bar{\bPsi} \bu_j  , \ \forall j = 1, ..., S, $ and $\lrb{  \bu_j }_{j=1}^S $ denotes the elements of $\calS_\rho$.    
Note that this last problem falls under the category of special LPs, and following the discussion in Definition~\ref{sec:app_specLP}, its solution is an elementary vector.  
Thus, the optimal solution to $(P_5)$ is given by $ \bx^{\star} =  \bu_{j^\star} , \ \textrm{where }  j^\star =  \argmin_{1 \leq j \leq S } \ \alpha_j $. 
Consequently, for the two antenna domain case, solving for $\bx^\star$ reduces to just finding the minimum of the $S$-dimensional vector, $\balp$. Although this is similar in complexity to exhaustively searching for $(P)$, it does provide a systematic means of doing that. 

\begin{lemma} \label{lem:opt_DWD}
Let $p(\bZ)$ be a non-convex function, and consider the following integer program
\begin{align}
(Q) \ \ \bZ^\star = \argmin \ p(\bZ)  \ \st \ \bZ \in \calS, ~~ \textrm{where}~ \calS = \lrb{\bW_j \ | \ j = 1, ..., n }  
\end{align}
Letting $[\bthe]_j \triangleq p(\bW_j) ,  \ j = 1, ...,n $, $(Q$) is equivalent to,
\begin{align}
(Q) \ \ \bt^\star =
\begin{cases}
  \argmin \ p_{d}(\bt) = \bt^T \bthe \ \ \ \st \ \bt^T \bone = 1, \ \bt \geq \bzer, 
\end{cases}
\end{align}
\end{lemma}
The proof follows from considering the following ``DW-like'' mapping, $\calS =\lrb{ \bZ = \sum_j t_j \bW_j \ | \  \bt^T \bone = 1, \ \bt \in \IB^n  } \ (g.1) $. Then, the cost in $(Q)$ is written as $ p(\bZ) = \sum_j t_j p(\bW_j)$.
Letting $\bt = [t_1, ..., t_n ]^T , $ and $ \theta_j =  p(\bW_j)$, 
$(Q)$ is equivalent to, 
\begin{align}
(Q) \ \argmin \ p_d(\bt) =  \bt^T \bthe  \ \ \st \ \bt^T \bone = 1, \ \bt \in \IB^n  
\end{align}
It can be verified that the mapping in $(g.1)$ is \emph{one-to-one} from $Z$ to $\bt$. 

\addcontentsline{toc}{chapter}{Bibliography}
\bibliography{ref_hadi_merged}
\bibliographystyle{ieeetr}

\end{document}